\documentclass[reprint, amsmath, amssymb, aps, prx, superscriptaddress, showpacs, floatfix]{revtex4-2}
\usepackage{natbib}
\usepackage{dcolumn}
\usepackage{bm}
\usepackage[normalem]{ulem} 
\usepackage{float}
\usepackage[utf8]{inputenc}
\usepackage{graphicx}
\usepackage{siunitx}
\usepackage{amsmath}
\usepackage{amssymb}
\usepackage{braket}
\usepackage{xcolor}
\usepackage{multirow}

\begin{document}
\title{
Hardware requirements for realizing a quantum advantage with deterministic single-photon sources
}

\author{Patrik I. Sund}
\affiliation{%
 Center for Hybrid Quantum Networks (Hy-Q), Niels Bohr Institute\\
 University of Copenhagen, Blegdamsvej 17, DK-2100 Copenhagen, Denmark
}%
\author{Ravitej Uppu}
\affiliation{%
 Center for Hybrid Quantum Networks (Hy-Q), Niels Bohr Institute\\
 University of Copenhagen, Blegdamsvej 17, DK-2100 Copenhagen, Denmark
}%
\affiliation{%
Department of Physics \& Astronomy, University of Iowa,
Iowa City, IA 52242
United States
}%
\author{Stefano Paesani}
\affiliation{%
 Center for Hybrid Quantum Networks (Hy-Q), Niels Bohr Institute\\
 University of Copenhagen, Blegdamsvej 17, DK-2100 Copenhagen, Denmark
}
\affiliation{NNF Quantum Computing Programme, Niels Bohr Institute, University of Copenhagen, Blegdamsvej 17, Copenhagen 2100, Denmark.}
\author{Peter Lodahl}
\email{lodahl@nbi.ku.dk}
\affiliation{%
 Center for Hybrid Quantum Networks (Hy-Q), Niels Bohr Institute\\
 University of Copenhagen, Blegdamsvej 17, DK-2100 Copenhagen, Denmark
}%

\date{October 16, 2023}

\begin{abstract}
Boson sampling is a specialised algorithm native to the quantum photonic platform developed for near-term demonstrations of quantum advantage over classical computers. 
While clear useful applications for such near-term pre-fault-tolerance devices are not currently known, reaching a quantum advantage regime serves as a useful benchmark for the hardware.
Here, we analyse and detail hardware requirements needed to reach quantum advantage with deterministic quantum emitters, a promising platform for photonic quantum computing. 
We elucidate key steps that can be taken in experiments to overcome practical constraints and establish quantitative hardware-level requirements. 
We find that quantum advantage is within reach using quantum emitters with an efficiency of $60\%-70\%$ and interferometers constructed according to a hybrid-mode-encoding architecture, constituted of Mach--Zehnder interferometers with an insertion loss of $\SI{0.0035}{\decibel}$ (a transmittance of $99.92\%$) per component.
\end{abstract}

\maketitle
Devices based on quantum systems can potentially outperform the capabilities of classical computers \cite{preskill2012quantum, preskill2018quantum}.
Quantum technologies are rapidly progressing towards this goal and new computational regimes are being explored~\cite{boixo2018characterizing, madsen2022quantum, morvan2023phase}.
While fault-tolerance is generally thought to be necessary to reach most practical applications, reaching this regime necessitates hardware requirements that are far from current capabilities, limiting demonstrations to small-scale experiments~\cite{maring2023generalpurpose, google2020hartree}.   
Quantum advantage (QA), where specialized algorithms can demonstrate speed-ups over classical computers, has been identified as an intermediate milestone computational regime amenable for near-term hardware using readily available quantum hardware components~\cite{aaronson2011computational, hamilton2017gaussian, boixo2018characterizing}.
While it is currently not known if any practical applications are possible in this regime, it serves as an entry point to beyond-classical capabilities and an important benchmark for developing scalable platforms that can evolve towards fault-tolerance.
In this context, we analyze the hardware requirements for achieving QA using photonic quantum hardware, where fusion-based approaches for fault-tolerant quantum computing have been proposed \cite{bartolucci2023fusion, Paesani2023}.

Aaronson and Arkhipov proposed boson sampling of photonic quantum states \cite{aaronson2011computational} as a route for demonstrating QA with near-term quantum hardware. 
The key insight is the connection between the correlations induced by linear interference operations on $N$ indistinguishable photons in an $M$-mode linear optical interferometer and the matrix permanent, a quantity that is $\#$P-hard to compute on a classical machine \cite{valiant1979complexity, Scheel2004}.
However, imperfections in photon sources and losses in optical interferometer networks and detectors rapidly diminish the degree of quantum correlations and overthrow the quantum advantage.
This loss of QA has been captured in models on `noisy' boson samplers that proposed efficient classical computation algorithms, thereby imposing strict bounds on the indistinguishability of photons and the overall optical loss \cite{brod2019photonic, renema2018classical, renema2018efficient}.
As quantum photonic hardware continues to rapidly advance  \cite{Elshaari2020np,uppu2021natnano,Pelucchi2021,Moody2022}, the formulation of quantitative benchmarks for realizing QA is a critical need as they provide milestones for guiding the hardware development. 

In this paper, we present a comprehensive analysis quantifying the performance metrics of the constituent building blocks essential for surpassing efficient classical algorithms with boson sampling.
By benchmarking our framework against state-of-the-art single-photon sources and photonic integrated circuits, we identify a realistic regime for conducting QA experiments.
Our analysis focuses on the case of boson samplers based on single-photon sources (discrete variable photonic qubits), where several proof-of-concept experiments have been carried out to date \cite{broome2013photonic, spring2013boson, crespi2013integrated, tillmann2013experimental, wang2019boson}.

Recent advances in deterministic photon sources employing semiconductor quantum dots have demonstrated the generation of $>$100 nearly-identical photons \cite{uppu2020scalable, tomm2021nn}, setting the stage for scaling up from proof-of-concept experiments.
A remaining challenge is to realize large photonic circuits with sufficiently low loss such that the large photonic resource can be processed and measured to demonstrate QA \cite{brod2019photonic, Pelucchi2021}. As the optical loss in an interferometer circuit is highly dependent on its design, i.e., the spatial arrangement of Mach-Zehnder interferometers (MZI), we analyze the requirements of the individual MZIs and their integration into an optimal architecture.
Combining the analysis of the source imperfections and optical circuit loss, we identify two key optimizations, a rectangular circuit architecture and the encoding of modes in multiple degrees of freedom, that could enable an unequivocal demonstration of QA.
We determine that an insertion loss of $~\SI{3.5}{\milli dB}$ per MZI interferometer, i.e., a transmittance of $\SI{99.92}{\percent}$ is sufficient for the optimal architecture using state-of-the-art photon sources and detectors. 
This sets a clear target metric for ongoing advances in photonic integrated circuits \cite{stojanovic2018monolithic, bao2023very}, and is already within reach for the specialized, fixed circuits employed in Refs. \cite{wang2019boson, zhong2020quantum}.

\section{The boson sampling algorithm and its validation}
\begin{figure}
    \includegraphics[width=0.98\columnwidth]{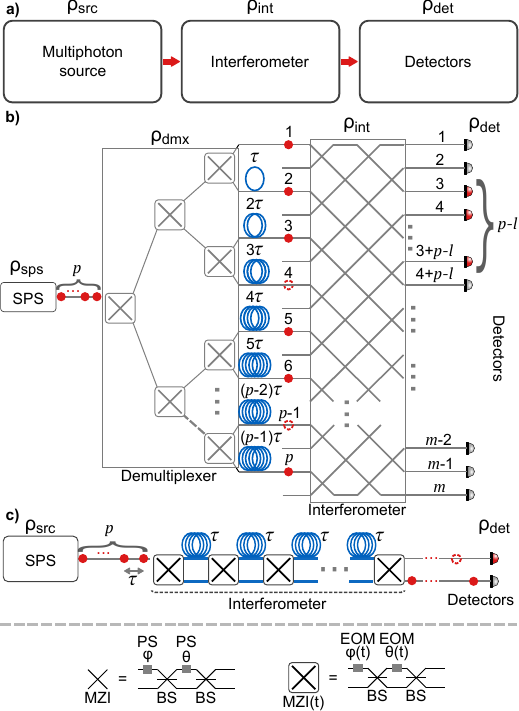}
    \caption{
   \textbf{a} A general boson sampling setup, consisting of a source of multi-photon input states, a multi-port interferometer, and detectors. The associated system losses $\rho$ are indicated for each sub-component. \textbf{b} A boson sampling setup based on a deterministic single-photon source (SPS) and a spatially encoded multimode interferometer. The single-photon source emits a string of single photons in pre-determined time bins, illustrated as filled red circles. A demultiplexer converts the photon stream into a $p$-photon source by deterministically switching each photon into a separate mode. The spatially encoded input state is sent into an interferometer constructed from a network of MZIs. Each MZI is illustrated as a cross, where each cross consists of two phase shifters (PSs) and two beamsplitters (BSs) arranged according to the illustration at the bottom. 
   \textbf{c} A boson sampling setup based on a deterministic single-photon source (SPS) and a time-bin interferometer. The SPS is identical to the case in (b) where $\tau$ denotes the time separation between subsequent photons. The interferometer consists of reconfigurable time-dependent MZIs, where time-dependent phase shifters are implemented using electro-optic modulators (EOMs), which are connected by delay lines illustrated as blue lines, where one loop corresponds to a single time-bin $\tau$ of delay. For both \textbf{b} and \textbf{c}, it is assumed that up to $l$ photons may be lost throughout the setup---illustrated as a red circle with a dashed outline---due to residual optical loss associated with each component.
    }
    \label{fig:scheme}
\end{figure}

Boson sampling is the task of sampling from the output photon distribution after multiple independent photons have interfered in a multimode linear optical interferometer.
The setup for implementing boson sampling is schematically illustrated in Fig.~\ref{fig:scheme}a) highlighting the key components: an input consisting of multiple indistinguishable photons, a large multimode interferometer module, and single-photon detectors.

The computational complexity of simulating boson sampling arises from the connection between multi-photon correlations and the calculation of matrix permanents. 
However, the computational hardness of calculating the matrix permanent decreases when duplicate rows or columns occur from multiple photons occupying the same input or output mode. 
To preserve the computational complexity, it is crucial to ensure collision-free states at both the inputs and outputs \cite{aaronson2011computational}, meaning that each mode contains at most one photon.  
Collision-free input states can be guaranteed by choosing the initial condition, where no more than one photon is injected into each input mode.
Ensuring collision-free outputs demands the interferometer to possess a large number of modes per photon. Specifically, the number of modes $m$ must scale at least quadratically with the number of photons $p$, i.e., $m \propto p^2$, a requirement arising from a phenomenon called the \textit{bosonic birthday paradox} \cite{aaronson2011computational, arkhipov2012bosonic}.
In experiments, collision-free outputs can be ensured through post-selection of events where photons are detected in the same number of output modes as input modes while discarding all other events.
This post-selection strategy remains applicable even when the detectors lack number-resolving capabilities, thereby enabling near-term implementations of QA with efficient single-photon detectors \cite{wang2019boson, zhong2020quantum, Bulmer2022}. 

To demonstrate QA, it is essential that the output samples can be validated as being computationally hard to produce by classical means \cite{brod2019photonic}. 
Due to the computational hardness, direct validation of the samples by comparison with exact distributions is infeasible. 
Within these constraints, the validation of QA through boson sampling requires two steps.
First, we must require that deviations in the experimental setup are small enough that approximate classical algorithms cannot simulate the output distribution efficiently.
Secondly, instead of validating that the samples are produced from the exact distribution, one verifies that the outputs are not reproduced by a computationally efficient distribution \cite{brod2019photonic, wang2019boson}. 
Specifically, statistical tests performed on the output samples obtained in a boson sampling experiment verify that the experimentally observed distribution differs significantly from a set of efficiently computable distributions.
These statistical tests provide a termination condition for the experiment, whereby the boson sampler is run until a sufficient number of output samples are generated to establish the statistical tests' convergence unequivocally. 
Thus, a QA demonstration will be feasible if a sufficient number of samples can be produced over an experimentally viable integration time, proportional to the sample acquisition rate, $r_{\text{sample}}$. The sample acquisition rate in an experiment is equal to the product of the generation rate of the multiple photon input state $r_{\text{input}}$ and the probability of the state reaching the detectors,  $P_{\text{sample}}$:
\begin{align}
\label{eq:sample_rate}
    r_{\text{sample}} = P_{\text{sample}} \cdot r_{\text{input}}.
\end{align}
The probability of sampling a $p$-photon coincidence at the output is related to the total per-photon efficiency of the system $P_{\text{sys}}$ as 
\begin{align}
\label{eq:sample_efficiency}
    P_{\text{sample}} = P_{\text{sys}}^p.
\end{align}
For convenience, we express these probabilities in decibels, i.e., $\rho_{\text{i}}=-10\log_{10}(P_{\text{i}})$. 
For brevity, we refer to the decibel probability, $\rho_{\text{i}}$ as loss, while $P_{\text{i}}$ is referred to as efficiency. 

\section{Experimental setup and imperfections}
As illustrated in Fig.~\ref{fig:scheme}a), the loss in the boson sampling architecture can be broken down into component-level losses as 
\begin{align}
\label{eq:system_loss}
    \rho_{\text{sys}} = \rho_{\text{src}} + \rho_{\text{int}} + \rho_{\text{det}},
\end{align}
where $\rho_{\text{src}}$ is the source loss,  $\rho_{\text{int}}$ is the interferometer loss, and $\rho_{\text{det}}$ is the detector loss. In this section, we detail the implementation and requirements on the components and discuss strategies to tackle experimental limitations in each component.
\subsection{Input state preparation}
The input state in the boson sampling algorithm consists of multiple indistinguishable single photons, each generated and encoded separately. 
Below, we discuss these two steps and any imperfections introduced in real experiments.

An important figure of merit for the photons employed in a boson sampling experiment is the pair-wise indistinguishability, $x^2$, defined as the overlap integral of photons with wavefunctions $\psi$ and $\psi'$, $x^2 = \braket{\psi'|\psi} = \int \psi'^*(t) \psi(t) \mathrm{d}t$. 
In experiments, the indistinguishability is quantified through the Hong--Ou--Mandel visibility in an interference experiment \cite{hong1987measurement}.
To ensure computational hardness, photon indistinguishability approaching near-unity visibility is necessary. 

Highly indistinguishable photons have been generated employing two approaches: 1) non-linear optics and 2) single quantum emitters. 
The former exploits the energy-time correlations in optical processes in nonlinear media like spontaneous parametric down-conversion and spontaneous four-wave mixing to generate correlated photon pairs. 
The resulting squeezed state can be used directly for Gaussian Boson Sampling \cite{hamilton2017gaussian, zhong2020quantum, madsen2022quantum}, or alternatively, the detection of one photon in the pair can be used to herald the presence of the other.
Despite being probabilistic, the heralded nature can be exploited through active feed-forward and multiplexing to realize a near-deterministic photon source \cite{migdall2002tailoring}. As the sources suffer from an intrinsic trade-off between the photon number purity (the probability of having one and only one photon pair per pulse) and the photon pair generation rate, this requires massive multiplexing of many probabilistic sources to reach near-deterministic operation with high number purity, which is an active area of research \cite{bartolucci2021switch}. 
These challenges and trade-offs could be overcome by leveraging the deterministic light-matter interactions enabled by coupling a highly-efficient single-photon emitter (e.g., a semiconductor quantum dot) to a nanophotonic structure \cite{lodahl2022deterministic}. 
Single-photon sources based on such deterministic light--matter interfaces have recently enabled the on-demand generation of $>$100 highly indistinguishable single photons operational at a rate of up to GHz, thus highlighting an avenue for realizing boson sampling in the QA regime \cite{uppu2020scalable}. 

Photons generated using such deterministic light-matter interfaces are naturally encoded in different time bins as determined by the excitation process of the quantum emitter.
While these time-bin encoded photonic states can be readily employed for boson sampling \cite{motes2014scalable, he2017time}, compatibility with photonic integrated circuits requires converting the time-bin encoded photons to spatial encoding using a demultiplexer \cite{wang2019boson, hummel2019efficient}. 
The demultiplexer takes an incoming stream of $p$ single photons separated in time and converts it to $p$ simultaneous photons in separate spatial modes.
In both time-bin (see Fig.~\ref{fig:scheme}c)) as well as demultiplexed spatial-bin encoding (Fig.~\ref{fig:scheme}b), in the demultiplexed spatial-bin encoded case, a single-photon source emitting at a rate of $r_{\text{single-photon}}$ is converted to a source of $p$ single photons emitted at a rate of $r_{\text{single-photon}}/p$. With a time-bin encoding, the rate will depend on the time-bin interferometer architecture employed, but for the case we will analyze, the input state generation rate will be $2r_{\text{single-photon}}/m$, where $m$ is the number of modes, such that $m/2$ is the number of time-bins, as will be explained in more detail later.

As quantum dot sources have proven capable of producing long strings of highly indistinguishable photons \cite{uppu2020scalable}, the scalability of this approach is mainly determined by the overall efficiency of generating single photons and delivering them to the demultiplexer,  as well as the actual efficiency of the demultiplexer setup.
We specifically consider and benchmark the case of a single deterministic source but note that access to multiple deterministic quantum dot sources would allow for spatially encoded experiments with higher input state generation rates and lower demultiplexer losses.  
The simultaneous use of multiple sources relies on the development of local tuning methods for overcoming intrinsic inhomogeneities of quantum dot sources, and important progress has recently been reported both for quantum dots in bulk samples \cite{zhai2022quantum} and in nanophotonic waveguides \cite{papon2022independent}.

A time-to-spatial mode demultiplexer can be realized by sending the emitted photon stream through a binary tree of switches. 
Each step in the tree doubles the number of spatial modes, as illustrated in Fig.~\ref{fig:scheme}b), such that the full demultiplexer requires a depth of $\lceil \log_2(p) \rceil$.
Each output mode from the demultiplexer requires a specific optical delay to synchronize all photons. 
We associate a loss of $\rho_{\text{switch}}$ with each switching operation, such that the overall loss of the demultiplexer is:
\begin{align*}
    \rho_{\text{dmx}} = \lceil \log_2(p) \rceil \rho_{\text{switch}}.
\end{align*}

Finally, there will be a coupling loss, $\rho_{\text{coupling}}$ associated with connecting the output of the demultiplexer to the input of the interferometer. Thus, the overall source loss is
\begin{align*}
    \rho_{\text{src}} = \rho_{\text{sps}} + \rho_{\text{dmx}} + \rho_{\text{coupling}},
\end{align*}
where $\rho_{\text{sps}}$ is the loss associated with the single-photon source itself, accommodating for inefficiencies associated with the generation of single photons and subsequent coupling from the cavity or waveguide applied \cite{lodahl2022deterministic}.

We note that the number of photons generated should ideally be kept as low as possible (for the targeted computational complexity) while maintaining intractability, as the sample acquisition rate will decrease exponentially with system efficiency for an increasing number of input photons in accordance with Eq. \eqref{eq:sample_efficiency}. 
In order to increase acquisition rates, experimental efforts typically employ a related algorithm called \textit{Aaronson--Brod} boson sampling \cite{aaronson2016bosonsampling}, where an additional $l$ photons are added to the $p$ input photons, while the outputs are post-selected to contain the same number of photons as before.
The probability of detecting the correct number of photons, i.e. the probability of generating a sample, $P_{\text{sample}}(p, l)$ can then be expressed as
\begin{align}
\label{eq:loss quadratic modes}
    P_{\text{sample}}(p, l) &= P_{\text{sys}}^{p-l} (1-P_{\text{sys}})^l \binom{p}{l},
\end{align}
where the factor $\binom{p}{l}$ is the number of combinations in which one can lose $l$ photons from $p$ input photons.
This leads to a speed-up in the sample acquisition rate, which increases combinatorially with the number of lost photons $l$. 
The downside is that post-selection increases the deviation from the ideal case and lowers the computational complexity. 
In practice, the simulation algorithm in Ref. \cite{renema2018classical} can approximate the output of such an experiment within an error $E$ bounded by
\begin{align}
\label{eq:approx_error}
    \frac{x^{2(k+1)}(\frac{p-l}{p})^{k+1}}{1-(x^2\frac{p-l}{p})} &\geq E^2,
\end{align}
where $x^2$ is the indistinguishability of the photons. Thus, the number of lost photons allowed depends on the indistinguishability of the photons in the input state. 

\subsection{Interferometer design and operation}
Multimode interferometers are typically constructed using cascaded arrays of Mach-Zehnder Interferometers (MZIs) and implement an $m$-dimensional unitary matrix operation, with the circuit accommodating $m$ input modes and $m$ output modes.
To validate boson sampling through statistical sampling of random unitary circuits \cite{aaronson2011computational}, experiments require either several static circuits or a programmable circuit of MZIs with adjustable beam splitters and phase shifters.
The latter approach is favored, as it offers access to a substantial number of random unitary operations \cite{carolanScience}. 
Universal interferometer architectures enable the implementation of all unitary transformations on $m$ modes, effectively generating any $m \times m$ unitary matrix \cite{reck1994experimental, clements2016optimal}. 

The per-photon loss for an interferometer architecture depends on the number of optical components, i.e. MZIs, a photon passes through from the input to the output. 
We will refer to this number of MZIs as the \textit{optical depth} $D(m)$, such that the interferometer loss can be written as
\begin{align*}
    \rho_{\text{int}} &= D(m) \cdot \rho_{\text{MZI}}.
\end{align*}
The depth will increase for a higher number of modes where the exact dependence is given by the specific architecture employed. 
As such, there are two main strategies that are employed to reduce loss: 1. Reducing the number of modes, and 2. Employing interferometer architectures with a lower optical depth for a given number of modes. 

\subsection{Reducing the number of modes}
Due to the challenges in scaling up low-loss interferometers, experiments involving a large number of photons ($p>10$) have employed interferometers with the number of modes $m$ smaller than $p^2$.
Consequently, this choice yields a proportionally reduced depth $D(m)$.
Although this reduction in the number of modes mitigates optical loss within the interferometer, the bosonic birthday paradox no longer holds, i.e., the outputs cannot be assumed to be collision-free. In this scenario, the output event where multiple photons occupy the same mode will be indistinguishable from photon loss unless detectors with number-resolving capabilities are utilized. However, using the framework established in Ref.~\cite{chin2018generalized}, it can be shown that the computational complexity of boson sampling with collisions is at least as high as the computational complexity of collision-free boson sampling with the same number of nonzero elements. For more details, we refer to Appendix A.

Assuming a post-selection criterion where the detection of photons is restricted to $d$ $(<p)$ modes, with $p$ representing the number of input photons, Eq. \eqref{eq:loss quadratic modes} can be reformulated to derive the total sample efficiency $P_{\text{sample, lin}}$ for a linear number of modes ($m\propto p$)
\begin{align}
\label{eq:linear_sampling_efficiency}
    P_{\text{sample, lin}}(p,d) &= \sum_{l=0}^{d}  P_{\text{sys}}^{(p-l)} (1 - P_{\text{sys}})^{l} \binom{p}{l} P_{\text{ps}}(p, d, l, m).
\end{align}
Here, $P_{\text{ps}}(p, d, l, m)$ is the probability of detecting an output state with $d-l$ collisions and $p-l$ photons from an interferometer with $m$ modes--essentially quantifying the effective post-selection efficiency.
This equation involves summing the probabilities of all possible collision and photon loss configurations that result in photodetection in $d$ modes.
These probabilities are then multiplied by the occurrence probability of the given combination of losses and collisions. 
The computation of Eq. \eqref{eq:linear_sampling_efficiency} relies on the knowledge of the effective post-selection efficiency $P_{\text{ps}}(p, d, l, m)$. 
To estimate this quantity, we assume uniform sampling of Haar-random scattering matrices in the Hilbert space. 
Thus, we can estimate the effective post-selection efficiency as the ratio between the size of the Hilbert space with $p-l$ photons in $m$ modes with $p-l-d$ collisions---essentially the post-selected portion of the Hilbert space---and the full Hilbert space for $p-l$ photons in $m$ modes. 
This is expressed as:
\begin{align}
    P_{\text{ps}}(p, d, l, m) = \frac{ \binom{m}{d} \cdot \binom{p-l-1}{p-l-d}}{\binom{m + p-l - 1}{p-l}. }
    \label{eq:effective_post_selection_efficiency},
\end{align}
where $\binom{a}{b}$ represents the binomial coefficient for $a$ and $b$. For more details, we refer to Appendix B.

\subsection{Path-encoded boson sampling circuit architectures}
\begin{figure}
    \includegraphics{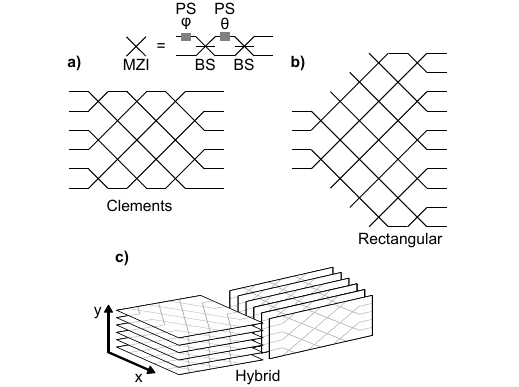}
    \caption{
    \textbf{a} Illustration of the Clements interferometer architecture, as detailed in Ref.~\cite{clements2016optimal}. Each cross corresponds to an MZI, which can be constructed from two 50:50 beamsplitters (BSs) and two phase-shifters (PSs). \textbf{b} Illustration of a rectangular interferometer with a larger number of output modes than input modes, the latter of which is equal to the number of input photons. The interferometer can be described by a rectangular matrix, hence the name. \textbf{c} Illustration of an interferometer with multiple mode-encodings, i.e. a hybrid mode-encoding. In this case, one mode encoding has spatial modes separated in the x-direction, while the other has modes separated in the y-direction.
    }
    \label{fig:architectures}
\end{figure}

The physical design of the interferometer architecture depends on the mode encoding employed. 
We will first present the architectures for spatial encoding. 
While theoretical works typically assume the use of universal interferometer architectures \cite{aaronson2011computational}, larger-scale experimental endeavors have so far featured interferometers constructed from non-universal architectures \cite{wang2019boson, zhong2020quantum, madsen2022quantum} with lower optical depth. Formally, the complexity arguments from Refs.~\cite{aaronson2011computational, aaronson2016bosonsampling} are valid only in the universal case, but practically, quantum advantage experiments are hard to simulate even in nonuniversal cases.
To provide an overview of how the different approaches compare, we will examine three interferometer architectures: a universal architecture, a fully-connected architecture with single mode-encoding, and a fully-connected hybrid-mode-encoded architecture. 
Here, ``fully connected" signifies that all of the output ports of the interferometer are connected to all of the input ports, generally resulting in a unitary matrix where all elements are non-zero, but that in general can be non-universal. 
In all three cases, the architectures were chosen to balance the loss per photon to the best possible degree for all input--output configurations. 
In the Clements case, and to a lesser extent the Rectangular case, paths along the edges traverse fewer MZIs.
Since the number of such edge cases is negligible, we assume uniform loss for all input--output configurations in our analysis.

The universal architecture introduced by Clements et al. \cite{clements2016optimal}, which we will refer to as the `Clements' architecture for simplicity, is constructed from $m$ columns of MZIs, as shown in Fig.~\ref{fig:scheme}a).
The interferometer loss in the Clements architecture is given by $\rho_{\text{int}} = m \cdot \rho_{\text{MZI}}$.
The interferometer loss can be optimized by employing non-universal architectures. 
We propose a non-universal `Rectangular' interferometer architecture (see Fig.~\ref{fig:scheme}b), which maintains full connectivity but reduces the depth $D$. This is achieved by reducing the number of input modes to be equal to the number of input photons $p$, while maintaining the same number of output modes $m$.
The interferometer consists of an initial section where two modes are added at the edges of each additional column of MZIs, and a second section fully connecting every input mode to all output modes. In practice, the Rectangular architecture is equivalent to starting a Clements architecture partway in, distributing input modes starting from the middle, and removing unused MZIs.
The interferometer loss, $\rho_{\text{int}}$, given by the number of MZI columns multiplied by the MZI insertion loss, is  $ \big(\lceil\frac{m}{2}\rceil+\lceil\frac{p}{2}\rceil - 1\big) \cdot \rho_{\text{MZI}}$.
Notably, when the number of output modes is much larger than the number of input modes, i.e. the number of input photons, such that $m \gg p$, $\rho_{\text{int}}$ is approximately halved compared to the Clements architecture.

Next, we introduce a `hybrid mode interferometer architecture' inspired by recent experiments \cite{wang2019boson, zhong2020quantum,madsen2022quantum}. 
These interferometers encode modes over multiple degrees of freedom, e.g. path or polarization, resulting in a hybrid mode encoding.  
For instance, one degree of freedom might represent spatial modes separated in the x-direction, while another represents spatial modes separated in the y-direction as shown in Fig.~\ref{fig:architectures}c).
Concatenating fully connected interferometers in each direction results in an interferometer that is fully connected across all modes. 
The power of this approach lies in the way the number of modes and the depth scale with degrees of freedom. 
The total number of modes in the interferometer is equal to the product $m = \prod_i m_i,$ where $m_i$ is the number of modes encoded over the $i$th degree of freedom. 
The optical depth, however, is equal to the sum of individual optical depths $\rho_{\text{int}} = \sum_i D_i \cdot \rho_{\text{MZI}, i},$
where $D_i$ is the depth for the interferometer connecting all modes for the $i$th degree of freedom, and $\rho_{\text{MZI}, i}$ is the MZI insertion loss for the $i$th degree of freedom. 
As an example, if we encode modes over two degrees of freedom, with $m_x=\sqrt{m}$ modes in the $x$ direction and $m_y=\sqrt{m}$ modes $y$ direction, the total number of modes remains as $m_x \cdot m_y = m$. 
A fully connected interferometer can then be constructed from Clements interferometers over the $m_x$ modes followed by Clements interferometers over the $m_y$ modes, as illustrated in Fig.~\ref{fig:architectures}c). 
The total optical depth will then be $D_x + D_y = m_x + m_y = 2 \sqrt{m}$.  
This approach allows for efficient scaling of both modes and depth, and it's worth noting that the Clements interferometers in each mode encoding can be replaced with Rectangular interferometers to reduce depth further.

The optical depth scalings for the different architectures can be summarized as follows
\begin{align}
\label{eq:interferometer_loss}
    \rho_{\text{int}} = \Bigg\{
    \begin{array}{lr}
        m \cdot \rho_{\text{MZI}} & \text{Clements,}\\
        \big(\lceil\frac{m}{2}\rceil + \lceil\frac{p}{2}\rceil - 1\big) \cdot \rho_{\text{MZI}} & \text{Rectangular,}\\
        \sum_i D_i \cdot \rho_{\text{MZI}, i} & \text{Hybrid}.
    \end{array}
\end{align}

\subsection{Time-bin encoded interferometer architectures}

\begin{figure}
    \includegraphics[width=0.98\columnwidth]{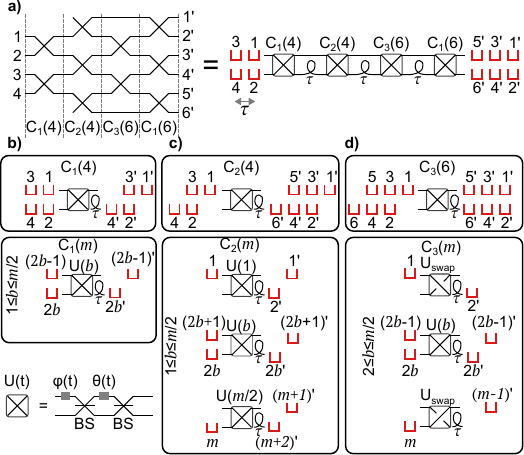}
    \caption{
    Illustration of how time-bin interferometers can be constructed to implement specific features in interferometer architectures. The red  bin symbols $\color{red}{\sqcup}$ correspond to the modes of the interferometer, and $\tau$ is the temporal separation between photons, which is equal to the inverse of the SPS emission rate $\tau = r^{-1}_{\text{single-photon}}$. Numbers correspond to input modes, and numbers with primes correspond to output modes. All of the time-bin interferometer architectures make use of reconfigurable MZIs, illustrated as boxed crosses. As is shown in the bottom-left corner, these MZIs are constructed using time-dependent phase shifters $\phi(\mathrm{t})$ and $\theta(\mathrm{t})$, which determine the unitary transformation U(t) effected by the interferometer. 
    \textbf{a} Illustration of a Rectangular interferometer architecture implemented with spatial modes on the left-hand side and with a time-bin interferometer on the right-hand side.
    The MZI columns are split into three categories $\text{C}_\text{1}(m)$, $\text{C}_\text{2}(m)$, and $\text{C}_\text{3}(m)$, where $m$ is the number of modes which is assumed to be an even number. \textbf{b} \textit{Upper} Illustration of a specific MZI column for four input modes, $\text{C}_\text{1}(4)$. \textit{Lower} Operation protocol of the $\text{C}_\text{1}(m)$ MZI column for a general even number of input modes, $m$. \textbf{c} \textit{Upper} Illustration of a specific MZI column for four input modes, $\text{C}_\text{2}(4)$. \textit{Lower} Operation protocol of the $\text{C}_\text{2}(m)$ MZI column for a general even number of input modes, $m$. This type of MZI column increases the total number of modes in the time-bin interferometer by two. \textbf{d} \textit{Upper} Illustration of a specific MZI column for six input modes, $\text{C}_\text{3}(6)$. \textit{Lower} Operation protocol of the $\text{C}_\text{3}(m)$ MZI column for a general even number of input modes, $m$. In the first and last time-bins, the MZI must be configured to perform the operation $\text{U}_{\text{swap}}$, which swaps the two input modes. This ensures that the number of output modes from the column is equal to the number of input modes $m$.}
    \label{fig:time-bin}
\end{figure}

Time-bin interferometers make use of time-dependent MZIs and fiber delays to implement multimode interferometers with significantly fewer physical resources than their spatially encoded counterparts. 
This is achieved by reconfiguring each physical MZI to implement a different transformation for every time-bin. 
Several interferometer architectures have been proposed and implemented \cite{motes2014scalable, he2017time, qi2018linear}, which differ in the number of spatial modes, and number of physical interferometers. 
We will analyze the architecture in Ref. \cite{he2017time}, with one modification -- instead of using a single MZI, we employ a cascaded series of multiple physical MZIs based on the interferometer employed in Ref. \cite{qi2018linear}. 
This choice is due to the lower total propagation loss and higher input generation rate of cascaded interferometers, as explained in Appendix C.
The multiphoton state input into this time-bin interferometer is encoded over two spatial modes of the MZI and $m/2$ time-bins. 
Each physical MZI in the cascade implements operations equivalent to a column of MZIs in the Clements or Rectangular interferometer, as illustrated in Fig.~\ref{fig:time-bin}.
The number of physical MZIs in the cascade then determines the depth $D(m)$ of the interferometer.
In addition, a relative time delay of one time-bin is introduced between the two output arms of the MZI between each step in the cascade, which ensures interference between photons in separate time-bins. 

Fig.~\ref{fig:time-bin} shows how certain spatial interferometers can be converted to time-bin interferometers. 
Specifically, we can construct the equivalent of spatial interferometers by combining three different MZI column types as shown in Fig.~\ref{fig:time-bin}a). 
All of the interferometer columns have the exact same construction for the time-bin interferometer architecture, a time-dependent MZI with a delay in one of the output modes, with the only difference being the sequence of transformations. 
The first MZI column type, labeled $\text{C}_1$ interferes all modes pairwise, starting with the first mode. 
The time-bin implementation of $\text{C}_1$ increases the total number of time-bins at the output by 1 due to the time delay, where the first mode and last mode will occupy their own time-bins. 
The second interferometer column type, $\text{C}_2$ differs from $\text{C}_1$ in the nature of the input state, where the first and last time-bins of the input states are occupied by only one mode, as shown in Fig.~\ref{fig:time-bin}c). 
The operation of the MZI on this asymmetric input state results in two additional modes (time-bins) at the output in addition to the time delay. 
The final interferometer column type, $\text{C}_3$, takes an input state where the first and final time-bins are only occupied by one of the modes and enacts a swap transformation
\begin{align*}
\text{U}_{\text{swap}} = \bigg[\begin{matrix}
0 & 1 \\
1 & 0
\end{matrix}\bigg],
\end{align*}
on the first and last time-bins, as shown in Fig.~\ref{fig:time-bin}d). 
This reduces the number of time bins by one compared to the input state. 
This type of column effectively interferes the modes pairwise starting with the second mode, such that the first and last modes don't interfere with any other mode.
 
In order to construct a specific architecture, we only need to identify the order and type of MZI columns that must be implemented.
A Clements interferometer can be implemented by combining $m/2$ pairs of $\text{C}_\text{1}(m)$ and $\text{C}_\text{3}(m)$ columns, such that the optical depth, i.e. the number of physical MZIs in the interferometer, is $m$. 
The Rectangular interferometer consists of a $\text{C}_\text{1}(2p)$ column followed by $(m-2p)/2$ $\text{C}_\text{2}(m')$ columns where $m'$ is increased from $2p$ at the input of the interferometer to $m$ at the interferometer by increasing the number of modes by 2 for every column. 
This is followed by $p-1$ pairs of $\text{C}_\text{3}(m)$ and $\text{C}_\text{1}(m)$ columns, resulting in an optical depth of $(m+2p)/2-1$. 
Note that we have increased the number of input modes from $p$ in Eq. \eqref{eq:interferometer_loss} to $2p$ as one SPS can only populate one of the spatial input modes as shown in Fig.~\ref{fig:scheme}, i.e. half of the input modes of the Rectangular interferometer.

To account for the effect of delay lines when establishing hardware requirements for time-bin interferometers, we can adjust the MZI insertion loss to include the propagation loss of the output with the longest delay
\begin{align}
    \rho_{\text{MZI, time-bin}} = \rho_{\text{MZI}} + \tau \cdot \rho_{\text{prop}} + \rho_{\text{coupling}},
\end{align}
where $\tau$ is the separation between time-bins, $\rho_{\text{prop}}$ is the propagation loss per unit time for the delay lines, and $\rho_{\text{coupling}}$ is the coupling loss associated with going from one MZI to the next, i.e. coupling into and out of delay lines. 

Lastly, we have to account for the empty time-bins that are required to implement an interferometer where the number of modes (output modes in the case of Rectangular interferometers) exceeds $2p$. 
The interferometer can only process $m/2$ time-bins at a time, which necessitates a temporal separation of $m/2$ time-bins between the start of two input states. 
Therefore, the input state generation rate, $r_{\text{input}}$, is limited in comparison to the single-photon generation rate as 
\begin{align}
    r_{\text{input}} = \frac{2r_{\text{single-photon}}}{m}.
\end{align}

\section{System benchmark for implementing the Aaronson--Brod boson sampling algorithm}
Before delving into the component-level hardware benchmarks, we address the overall system requirements, focusing on the two key system parameters, photon indistinguishability $x^2$ and system loss $\rho_\textrm{sys}$.
We analyze the interplay between these two parameters in implementing Aaronson--Brod's boson sampling algorithm with lost photons.
To provide a practical assessment of hardware performance, we choose realistic experimental conditions for run time and error rate of approximate classical algorithms for QA demonstrations with $p \geq 50$ photons.

We use the coupon collectors problem \cite{ferrante2012note, uppu2020scalable} to estimate the number of samples required for validation 
\begin{equation}
    r_{\text{sample}}\cdot t_{\text{integration}} \approx m \log(m) / p, 
\end{equation}
where $t_{\text{integration}}$ is the total run time of the boson sampler and ${r_{\text{sample}} \cdot t_{\text{integration}}}$ is the total number of samples acquired. 
We set the target for $r_{\text{sample}}$ to be 100 samples per day, $r_{\text{sample}} = \SI[quotient-mode=fraction,parse-numbers=false]{100/(24\cdot3600)}{\hertz}$, which is in line with the measured $\SI[quotient-mode=fraction,parse-numbers=false]{6/3600}{\hertz}$ photon coincidence rate reported in Ref.~\cite{wang2019boson} that allowed validating the boson sampling experiment. 

We analyze the boson sampling experiment for a deterministic single-photon source at the input that emits photons at a rate of 1 GHz. We consider both the case with a demultiplexer and a spatially encoded interferometer as well as the case with a time-bin interferometer.
In the former, the $p$ photon input state is generated at a rate of $r_{\text{input, spatial}} = 1/p\SI[quotient-mode=fraction,parse-numbers=false]{}{\giga\hertz}$. In the latter, the input state containing $M/2$ time-bins is generated at a rate of 
$r_{\text{input, time-bin}} = 2/m\SI[quotient-mode=fraction,parse-numbers=false]{}{\giga\hertz}$.
We consider two circuit architectures of the interferometer: one with a quadratic number of modes $m=(p-l)^2$ and another with a linear number of modes $m=10\cdot(p-l)$, as detailed in Sec. II.A and II.C, respectively. 
By comparing the two cases we can examine the influence of mode scaling on the trade-off between photon indistinguishability and system loss tolerance. 
We combine Eq. \eqref{eq:sample_rate} with Eq. \eqref{eq:loss quadratic modes} (Eq. \eqref{eq:linear_sampling_efficiency})for the quadratic (linear) case, to find the level of loss that results in an $r_{\text{sample}}$ of 100 samples per day. 

We choose an error rate of $E\leq0.01$, given by Eq. \eqref{eq:approx_error}, for approximating the noisy boson sampling output using classically computed permanents of order $k=49$. We plot this relation for a varying number of detected photons $(p-l)$, where the number of lost photons $l$ has been set as high as possible while keeping the error of the approximation below the threshold of $E\leq0.01$.

\begin{figure*}
    \includegraphics{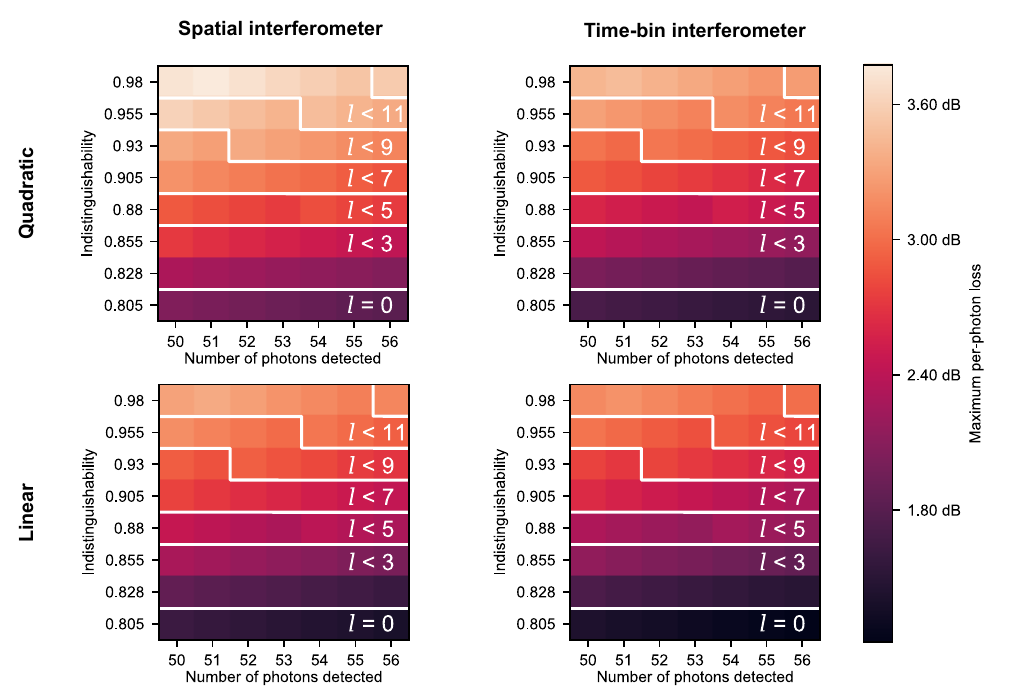}
    \caption{
    Maximum per-photon loss for the full circuit to perform boson sampling in the QA regime versus degree of indistinguishability and number of detected photons and for different values $l$ of lost or colliding photons. The two upper plots are for the case where the number of modes scales quadratically with the number of photons $m = (p-l)^2$, and the two bottom plots  are for the case where the number of modes scales linearly with the number of photons $m = 10 \cdot (p-l)$. The two plots to the left are for spatial interferometers with a demultiplexed source, whereas the two plots on the right are for time-bin interferometers. White contours indicate the added number of lost photons $l$, which increases with the indistinguishability, and detected number of photons.
    }
    \label{fig:tolerated_loss}
\end{figure*}

Results for mode scalings $m = (p-l)^2$ and $m = 10\cdot (p-l)$ are shown in the upper and lower rows of Fig.~\ref{fig:tolerated_loss}, respectively. 
The lowest indistinguishability $x^2 \approx 0.805$, found by setting $l=0$ and $p=50$ in Eq. \eqref{eq:approx_error}, corresponds to the maximal per-photon loss for Aaronson--Arkhipov boson sampling, i.e. with no photon loss or collisions. 
Increasing the photon indistinguishability allows for a higher number of lost photons with Aaronson--Brod sampling, increasing the per-photon loss that can be tolerated.
We find that $>$ 3 dB (i.e. $50 \%$) loss tolerance for realistic degrees of indistinguishability of quantum dot SPSs \cite{lodahl2022deterministic} for all four cases. We find the highest maximal  loss values at an indistinguishability of $x^2=0.98$ to be, from highest to lowest: $\SI{3.78}{\decibel}$ for the quadratic spatial case, $\SI{3.46}{\decibel}$ for the quadratic time-bin case, $\SI{3.35}{\decibel}$ for the linear spatial case, and $\SI{3.20}{\decibel}$ for the linear time-bin case. For comparison, in the limit of perfect indistinguishability, $x^2=1$ where up to $l=12$ photons can be lost with 50 detected photons, the highest maximal loss values would be $\SI{3.96}{\decibel}$ for the quadratic spatial case, $\SI{3.65}{\decibel}$ for the quadratic time-bin case, $\SI{3.53}{\decibel}$ for the linear spatial case, and $\SI{3.39}{\decibel}$ for the linear time-bin case. 

As the number of detected photons increases the total loss of the system typically increases, leading to a decreased maximum per-photon loss. However, for a fixed degree of indistinguishability, gradually increasing the number of detected photons can lead to abrupt changes when an additional lost photon can be tolerated according to Eq. \eqref{eq:approx_error}.
Thus, the optimum number of lost photons and the detected photons will both depend on the exact photon indistinguishability in the experiment.

In comparing the different plots it is evident that quadratic mode scalings and spatial interferometers lead to higher overall loss tolerance compared to the linear mode scaling and time-bin interferometers. The advantage of quadratic mode scaling can be attributed to the added effective post-selection loss associated with linear mode-scaling, described by Eq. \eqref{eq:effective_post_selection_efficiency}. Specifically, the average difference between quadratic and linear mode scalings is $\SI{0.430}{\decibel}$ for spatial interferometers and $\SI{0.277}{\decibel}$ for time-bin interferometers, in favor of quadratic interferometers. The advantage of spatial interferometers can be attributed to the lower input state generation rate for time-bin interferometers. The average difference between spatial and time-bin interferometers is $\SI{0.298}{\decibel}$ for a quadratic number of modes and $\SI{0.145}{\decibel}$ for a linear number of modes, both in favor of spatial interferometers.

Although quadratic mode-scalings allow for higher per-photon loss, the interferometers consist of more MZIs. As such, there is a trade-off between lower effective post-selection loss for quadratic interferometers and lower interferometer loss for linear mode scalings where the MZI insertion loss determines which mode scaling is favored.

Time-bin interferometers have a similar trade-off, as they have a lower maximal per-photon loss due to the lower input state generation rate, but do not require the use of a demultiplexer. However, a demultiplexer can be constructed from the same MZIs that are used to construct a time-bin interferometer, and as such, this trade-off can also be quantified in terms of MZI insertion loss. Specifically, a demultiplexer has an optical depth of $\lceil \log_2 (p) \rceil = 6$, where the equality holds for the optimal number of detected and lost photons for all indistinguishabilities considered in Fig.~\ref{fig:tolerated_loss}. The demultiplexer also involves a delay on all except the last photon, where the first photon has to go through the longest delay of $(p-1)$ time-bins. As for the time-bin interferometer, all of the $D(m)$ MZIs in the interferometer will in the worst case include one time-bin of delay which is not present in the spatial case. If we compare the added per-photon loss from the demultiplexer in the spatial case with the added delay and lower maximum per-photon loss in the time-bin case, we can find the following inequality for the regime where time-bin interferometers are less favorable implementations than spatial interferometers:
\begin{align}
\begin{split}
    6 \cdot \rho_{\text{MZI}} + (p-1) \rho_{\text{prop}} &\leq (D(m)-1) \rho_{\text{prop}} + \Delta, \\
    \Delta = \rho_{\text{sys, spatial}} &- \rho_{\text{sys, time-bin}}.
\end{split}
\end{align}
The average of the value for $\Delta$, i.e. the difference between maximal per-photon loss for spatial and time-bin interferometers, was found to be $\SI{0.145}{\decibel}$ for the case where $m=10(p-l)$ and $\SI{0.298}{\decibel}$ for the case where $m=(p-l)^2$. If we neglect propagation loss, $\rho_{\text{prop}} = 0$, and insert the average values we can estimate this inequality in the two cases considered in Fig.~\ref{fig:tolerated_loss}:
\begin{align}
    \begin{split}
        \rho_{\text{MZI}} &\leq \SI{0.05}{\decibel} \quad m = (p-l)^2, \\
        \rho_{\text{MZI}} &\leq \SI{0.024}{\decibel} \quad m = 10\cdot (p-l).
    \end{split}
    \label{eq:time_vs_space_inequalities}
\end{align}
In practice, coupling into the delay between MZIs will inevitably incur loss, and as such, these inequalities present a best-case scenario in favor of time-bin interferometers.

\section{Benchmarking hardware requirements}

The requirements on component losses for a given interferometer architecture can be found by combining Eqs. \eqref{eq:system_loss}, \eqref{eq:interferometer_loss}, with either Eq. \eqref{eq:loss quadratic modes} or Eq. \eqref{eq:linear_sampling_efficiency}.
To simplify the analysis, we note that only the interferometer loss scales with the number of modes, and separate the losses into the interferometer loss, $\rho_{\text{int}}$, and the remaining system loss, $\rho_{\text{sys}} - \rho_{\text{int}} = \rho_{\text{sps}} + \rho_{\text{dmx}} + \rho_{\text{coupling}}$.
We fix the degree of indistinguishability to $x^2 = 0.96$, which is readily achievable with present-day quantum dot single-photon sources \cite{ding2016demand}, while routes to achieve even higher values have been laid out \cite{Dree_en_2018}. This allows us to fix the number of input photons to $p=59$ with up to $l=9$ lost photons in accordance with Eq. \eqref{eq:approx_error}. 
We also consider the requirements for Aaronson--Arkhipov sampling where we post-select on detecting the same number of photons as are sent into the interferometer, i.e. fixing the number of input and output photons to $p=50$. The hardware requirements on the interferometer can be formulated as specific requirements on the MZI insertion loss by specifying the architecture and number of modes used for the interferometer.

\begin{figure*}
    \includegraphics{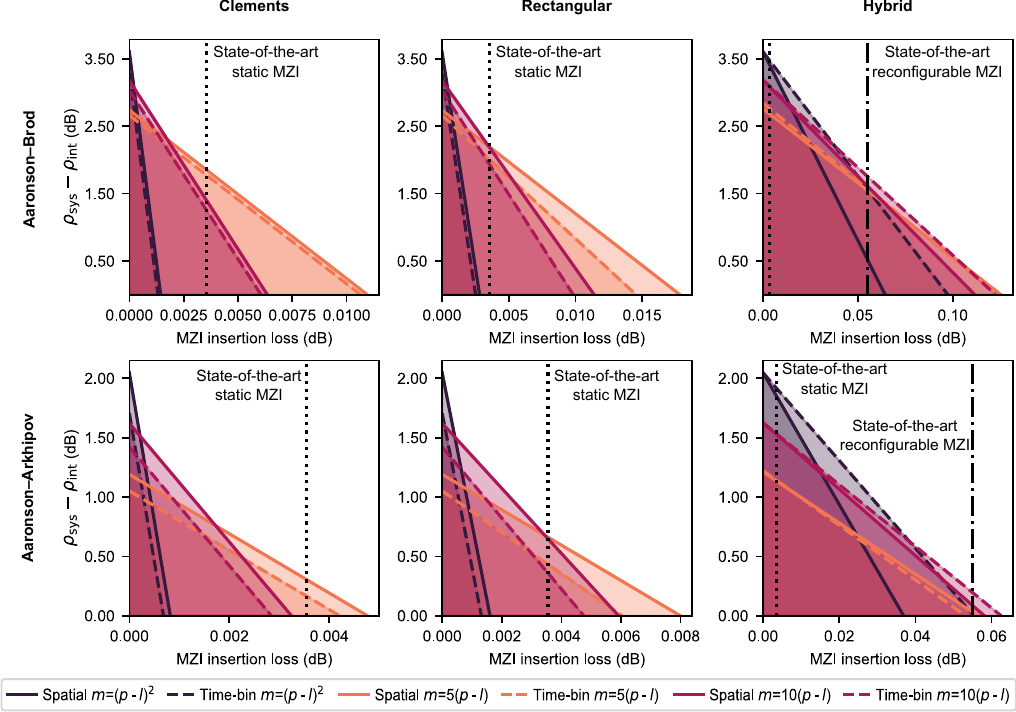}
    \caption{Plot of the requirements on MZI insertion loss (x-axis) and $\rho_{\text{sys}} - \rho_{\text{int}}$ (y-axis) with photon indistinguishability set to $x^2 = 0.96$. The upper plots show the requirements for Aaronson--Brod boson sampling, where the input state consists of $59$ photons with the outputs post-selected to contain $50$ photon detection events. The lower plots show the requirements for Aaronson--Arkhipov boson sampling, where we send in $50$ photons and detect $50$ photons, not allowing for photon loss or collisions. The first, second, and third column shows hardware requirements for setups where the interferometers are constructed according to the Clements architecture, the Rectangular architecture, and a set of Hybrid architectures, respectively. For the Clements and Rectangular architectures, the solid lines correspond to the requirements for spatially encoded architectures, whereas dashed lines correspond to the requirements for time-bin architectures. For the Hybrid architectures, the solid lines correspond to requirements for an interferometer encoded over two spatial mode-encodings, whereas dashed lines correspond to requirements for an interferometer encoded over time-bins and two spatial mode encodings. The dotted vertical lines mark the estimated MZI loss for a state-of-the-art experimental realization with static, nonprogrammable MZIs \cite{wang2019boson}. The dash-dotted vertical line in the plots for Hybrid interferometers marks the estimated MZI loss for a state-of-the-art experimental realization with programmable MZIs \cite{taballione202320}. } 
    \label{fig:src_vs_mzi}
\end{figure*}

\noindent \textit{Single mode-encoding:}
The first two columns of Fig.~\ref{fig:src_vs_mzi} show the requirements for three different choices of mode scaling for both Clements and Rectangular interferometer architectures, where the top figures show requirements for Aaronson--Brod \cite{aaronson2016bosonsampling} boson sampling, and the bottom figures show requirements for Aaronson--Arkhipov boson sampling \cite{aaronson2011computational}, i.e. with post-selection on the same number of detected photons as input photons. The figures include a vertical dotted line corresponding to the state-of-the-art insertion loss for a static MZI, estimated to be $-10\cdot\log_{10}(0.987)/(10+6)\:\SI{}{\decibel} \approx \SI{0.0035}{\decibel}$, which is the overall interferometer efficiency in Ref. \cite{wang2019boson} divided by the number of MZIs corresponding to the optical depth of a 10-mode interferometer followed by a 6-mode interferometer. 

Comparing the requirements on MZI insertion loss with the Eqs.~\eqref{eq:time_vs_space_inequalities}, we find that we are in the regime where demultiplexing and spatial mode encoded interferometers are favorable in terms of loss even with access to the rapidly reprogrammable MZIs required to construct a time-bin interferometer.

As seen from the figure, a QA demonstration using single-mode-encoding architectures would be within reach if one could use interferometers with state-of-the-art efficiency in conjunction with a source and detection efficiency of around $P_{\text{sps}}\cdot P_{\text{dmx}}\cdot P_{\text{coupling}} \cdot P_{\text{det}} \geq 0.65$.
This overall efficiency is currently beyond the state-of-the-art values reported with quantum dot sources\cite{maring2023generalpurpose, wang2019boson, chen2023heralded,wang2023deterministic}, see Table \ref{tab:literature_values} for an overview of parameters already reported in the literature. Further expected near-term improvements of the approach is also listed in the table, indicating that explicit QA demonstration with single-mode encoding is not far outside reach. 
\begin{table*}[t]
    \centering
    \begin{tabular}{c|c|c|c|c|c|c}
        Reference & $P_{\text{sps}}$ & $P_{\text{coupling}}$ & $P_{\text{dmx}}$ ($p$) & $P_{\text{det}}$ & Total efficiency & $x^2$ \\
        \hline
        Current best & 0.658 \cite{chen2023heralded} & 0.902 \cite{wang2019boson} & 0.83 (20) \cite{wang2019boson} & 0.95 \cite{madsen2022quantum} & 0.47 & 0.964 \cite{ding2016demand} \\
        Near-term estimated best & 0.78 \cite{uppu2020scalable} & 0.902 \cite{wang2019boson} & 0.92 \cite{uppu2020scalable} & 0.95 \cite{maring2023generalpurpose} & 0.615 & 0.985 \cite{ding2016demand, frejaThesis}
    \end{tabular}
    \caption{Table of state-of-the-art system efficiencies. In the column for demultiplexer efficiency, the number of modes of the demultiplexer employed is indicated in parentheses. For the current best indistinguishability, we have used the measured photon indistinguishability, whereas for the near-term estimated best we have used the estimated intrinsic photon indistinguishability which has been corrected for experimental imperfections.}
    \label{tab:literature_values}
\end{table*}

\noindent \textit{Hybrid mode encoding:}
Hybrid encoding schemes make the algorithms more robust to optical loss and hence put QA demonstrations within closer reach. We consider two distinct Hybrid architectures, one with two spatial mode encodings, like the one employed in Refs.~\cite{wang2019boson,zhong2020quantum} and illustrated in Fig.~\ref{fig:architectures}c), and one with two spatial mode encodings and one time-bin encoding. For the former, the number of modes in each encoding was chosen to optimize the optical depth, as described by Eq.~\eqref{eq:interferometer_loss}. This optimization procedure allowed for the total number of modes to be slightly increased if it led to favorable optical depth. Rectangular interferometers were employed in each mode encoding to minimize depth. For the latter encoding, we considered an architecture where the input state is partially demultiplexed, whereby sets of two spatial modes are inserted into a time-bin interferometer. By ensuring that there are no empty time-bins in the time-bin interferometer, we avoid the issue of a lowered input state generation rate for time-bin interferometers.  The output modes of the time-bin interferometers are then sent into a two-spatial-mode-encoding Hybrid interferometer employing Rectangular interferometers across the two spatial encodings. The total optical depth of this interferometer, including the initial demultiplexer is equal to:
\begin{align}
    D(n, p, m_1, m_2) = n + 2\lceil\frac{p}{2^n}\rceil + \lceil\frac{m_1}{2}\rceil + \lceil\frac{m_2}{2}\rceil +\frac{2^n}{2},
    \label{eq:depth_hybrid_time_bin}
\end{align}
where $n$ is the depth of the demultiplexer, such that there are $2^n$ spatial modes and $\lceil\frac{p}{2^n}\rceil$ time-bins after the demultiplexer, and where $m_1$ and $m_2$ correspond to the number of output modes in each spatial encoding. Similarly to the case of the spatial hybrid interferometer, we allow for the number of modes to be increased if it leads to a lower optical depth, only requiring that
\begin{align*}
    m_1\cdot m_2 \cdot \lceil\frac{p}{2^n}\rceil \geq m,
\end{align*}
where the left-hand side is the actual number of modes, and the right-hand side is the target number of modes, calculated from the mode scaling.
The depth of the demultiplexer and the number of spatial modes were optimized in order to minimize Eq.~\eqref{eq:depth_hybrid_time_bin}. For the case of a quadratic number of time-bins, the ideal demultiplexer depth was found to be $n=3$, with $m_1 = m_2 = 18$, whereas for the linear mode scalings, the ideal demultiplexer depth was found to be $n=4$ with the number of spatial modes equal to $m_1=11, \; m_2=12$ and $m_1=8, \; m_2=9$ for the case where $m=10(p-l)$ and $m=5(p-l)$, respectively.

The resulting hardware requirements for Aaronson--Brod (Aaronson--Arkhipov) boson sampling are shown in the top (bottom) plot of the right column of Fig.~\ref{fig:src_vs_mzi}, where the solid lines (`Spatial') refer to the case with two spatial mode encodings, and the dashed lines (`Time-bin') refer to the case with time-bin and two spatial encodings. In addition to the dashed line for the state-of-the-art static MZI insertion loss, the plots include a dashed-dotted line marking the state-of-the-art insertion loss for a reconfigurable MZI, estimated from Ref.~\cite{taballione202320} to be $\SI{1.1}{\decibel}/20 = \SI{0.055}{\decibel}$, which is the interferometer insertion loss divided by the number of MZIs. 

The hybrid encoding with the time-bin encoding performs better at larger mode-scalings but seems to perform comparably or slightly worse for the case where $m=5(p-l)$. The advantage at higher mode-scalings comes from the fact that the number of modes is distributed over three encodings, which means that the sum of the number of modes can be smaller. This is less of an advantage for the case where $m=5(p-l)$, where the additional demultiplexer loss included in the time-bin case gives a higher overall depth. It should be noted that the spatial hybrid interferometer requires the addition of a demultiplexer at a depth of $n=\lceil \log_2(p) \rceil$, which should be included as part of the source efficiency. As such, the time-bin hybrid interferometer would be expected to perform advantageously even for low mode-scaling.

Figure~\ref{fig:src_vs_mzi} clearly shows that the MZI insertion loss determines whether a quadratic mode scaling or linear mode scaling is favorable. Specifically, at the state-of-the-art static MZI insertion loss, employing an interferometer with a quadratic mode-scaling is best, whereas linear mode scalings are favored when using MZIs with the state-of-the-art reconfigurable MZI insertion loss. This is not the case for Clements or Rectangular architectures except for the case where the MZI insertion loss is vanishingly low. The discrepancy between the Clements and Rectangular architectures and the Hybrid architecture is attributed to how the optical depth scales with the number of modes, as shown in Eq.~\eqref{eq:interferometer_loss}. With modes distributed across two mode-encodings, the optical depth of Hybrid interferometers scales with $\sim\sqrt{m}$, as opposed to linearly in $m$.  As increasing the number of modes has a smaller impact on interferometer loss, the added effective post-selection loss associated with linear mode-scalings has a proportionally higher impact on the system loss for hybrid interferometers.

\section{Hardware requirements for near-term QA demonstrations}

From Fig.~\ref{fig:src_vs_mzi}  it is observed that an explicit QA demonstration  is within reach with a deterministic quantum dot source. Indeed using state-of-the-art static interferometers would imply that QA is reached for $P_{\text{sps}}\cdot P_{\text{dmx}}\cdot P_{\text{coupling}} \cdot P_{\text{det}} \geq 0.45$, where the required efficiency of each sub-component was already realized experimentally, see Table~\ref{tab:literature_values}. As time-bin interferometers require the use of reprogrammable MZIs, one could not make use of hybrid architectures with time-bin encoding in this case. As for state-of-the-art reconfigurable interferometers, this would require a setup with combined source and detection efficiencies around $0.65$ ($0.7$) for the time-bin (spatial) Hybrid architectures. These values are reachable with the estimated near-term values of the approach, cf. Table~\ref{tab:literature_values}. It is important to note that the state-of-the-art values of MZIs hold for thermo-optic phase-shifters, which are unsuitable for realizing time-bin interferometers due to their slow response time. Consequently, this would limit the present implementations of Hybrid time-bin architectures. On the other hand, Hybrid interferometers with spatial mode encodings appear to be promising candidates for near-term QA demonstrations with quantum dot single-photon sources.

It is clear from Table~\ref{tab:literature_values} that the source efficiency is the main bottleneck in realizing a demonstration of QA. In the following, we restrict the focus to the exact requirements for the single-photon source by fixing other losses to state-of-the-art values.
We fix the MZI insertion loss to the state-of-the-art value for static MZIs shown in Fig.~\ref{fig:src_vs_mzi}. 
The demultiplexer efficiency and coupling losses are fixed to realistic parameters extrapolated from Refs. \cite{zhong2020quantum, uppu2020scalable}:
\begin{align*}
    \rho_{\text{dmx}} &= \frac{0.458}{5} \lceil \log_2 (p) \rceil\SI[quotient-mode=fraction,parse-numbers=false]{}{\decibel}, \\
    \rho_{\text{coupling}} &= \SI{0.458}{\decibel}.
\end{align*}

\begin{figure}
    \includegraphics{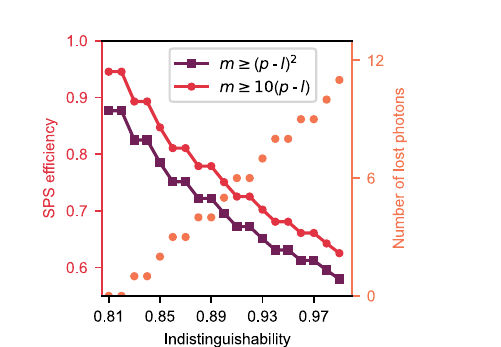}
    \caption{Requirements on source efficiency for a given indistinguishability with a hybrid interferometer encoded in two mode-encodings as per Fig. 1d). Rectangular interferometers have been used within both mode encodings. The requirements are defined to allow for 100 samples to be obtained per day with a 1 GHz single-photon generation rate. For each value of indistinguishability, the number of photons detected and lost has been optimized to increase the loss tolerance while maintaining an error bound higher than 1\% for the approximation algorithm. a) Quadratic mode scaling b) Linear mode scaling}
    \label{fig:indist_vs_src}
\end{figure}

We examine how the requirements on the source efficiency change as a function of the photon indistinguishability. The resulting curves for quadratic and linear mode-scalings are shown in Fig.~\ref{fig:indist_vs_src}. 
It is clear from the figure that quadratic mode-scaling is favored regardless of photon indistinguishability. For realistic photon indistinguishability, $x^2 \geq 0.96$, the results show that a QA demonstration is within reach for single-photon source efficiencies greater than $0.6$, which has been demonstrated experimentally \cite{chen2023heralded}. The challenge will be to construct a demultiplexer and an interferometer that are sufficiently large while maintaining sufficiently low loss, and connecting them to an exceedingly large number of low-loss detectors.

\section{Conclusion}
In conclusion, we have presented an in-depth analysis of the hardware requirements for realizing boson-sampling in the QA regime with deterministic single-photon sources, notably quantum dots in nanophotonic cavities and waveguides. The estimated benchmarks provide precise requirements on optical circuits and single-photon sources that must be reached, thereby offering a roadmap for future engineering efforts to realize that goal. Our analysis elucidates the precise advantages and disadvantages of strategies that are commonly employed in experiments to lower hardware requirements, such as making use of specialized interferometer architectures and employing interferometers with linear mode scaling. We have identified interferometers with hybrid mode encoding and quadratic mode scaling as a key strategy to demonstrating QA, an approach that has yet to see realization in experiments. Specifically, we have shown that a QA experiment based on single-photon boson sampling is within reach of current state-of-the-art hardware, provided that one can reach source efficiencies as high as 60\%-70\%.

In examining the requirements for time-bin encoded interferometers, we have found time-bin interferometers utilizing the Clements and Rectangular architectures to be inferior to equivalent spatial interferometers for QA demonstrations. This is due to the disadvantages associated with empty time-bins in the input state in the regime where MZI insertion loss is low enough for a QA demonstration to be feasible. For hybrid interferometer architectures, however, encoding a subset of the modes in time-bins leads to improved hardware requirements on MZI insertion loss. As such, developing rapidly reconfigurable photonic integrated circuits with bandwidth to support time-bin-compatible MZIs with sufficiently low loss is a promising direction to enable QA demonstrations.

Our analysis has focused on the requirements of the interferometers and single-photon sources, however, an underlying assumption for parts of the analysis was that the loss associated with coupling photons from the source and into the interferometer were comparable with the losses quoted in Refs.~\cite{wang2019boson, zhong2020quantum}. Achieving ultra-low-loss chip-to-fiber coupling is an important engineering challenge and an area of active research \cite{tiecke2015efficient, notaros2016ultra, marchetti2019coupling}. 
Ultimately coupling losses could be further mitigated by a partial or full-scale system integration, whereby sources, demultiplexer, interferometers, and detectors would be combined in a single device \cite{uppu2021natnano}, which constitutes an important future research direction.

\bibliography{biblio.bib}

\begin{thebibliography}{57}%
\makeatletter
\providecommand \@ifxundefined [1]{%
 \@ifx{#1\undefined}
}%
\providecommand \@ifnum [1]{%
 \ifnum #1\expandafter \@firstoftwo
 \else \expandafter \@secondoftwo
 \fi
}%
\providecommand \@ifx [1]{%
 \ifx #1\expandafter \@firstoftwo
 \else \expandafter \@secondoftwo
 \fi
}%
\providecommand \natexlab [1]{#1}%
\providecommand \enquote  [1]{``#1''}%
\providecommand \bibnamefont  [1]{#1}%
\providecommand \bibfnamefont [1]{#1}%
\providecommand \citenamefont [1]{#1}%
\providecommand \href@noop [0]{\@secondoftwo}%
\providecommand \href [0]{\begingroup \@sanitize@url \@href}%
\providecommand \@href[1]{\@@startlink{#1}\@@href}%
\providecommand \@@href[1]{\endgroup#1\@@endlink}%
\providecommand \@sanitize@url [0]{\catcode `\\12\catcode `\$12\catcode
  `\&12\catcode `\#12\catcode `\^12\catcode `\_12\catcode `\%12\relax}%
\providecommand \@@startlink[1]{}%
\providecommand \@@endlink[0]{}%
\providecommand \url  [0]{\begingroup\@sanitize@url \@url }%
\providecommand \@url [1]{\endgroup\@href {#1}{\urlprefix }}%
\providecommand \urlprefix  [0]{URL }%
\providecommand \Eprint [0]{\href }%
\providecommand \doibase [0]{https://doi.org/}%
\providecommand \selectlanguage [0]{\@gobble}%
\providecommand \bibinfo  [0]{\@secondoftwo}%
\providecommand \bibfield  [0]{\@secondoftwo}%
\providecommand \translation [1]{[#1]}%
\providecommand \BibitemOpen [0]{}%
\providecommand \bibitemStop [0]{}%
\providecommand \bibitemNoStop [0]{.\EOS\space}%
\providecommand \EOS [0]{\spacefactor3000\relax}%
\providecommand \BibitemShut  [1]{\csname bibitem#1\endcsname}%
\let\auto@bib@innerbib\@empty
\bibitem [{\citenamefont {Preskill}(2012)}]{preskill2012quantum}%
  \BibitemOpen
  \bibfield  {author} {\bibinfo {author} {\bibfnamefont {J.}~\bibnamefont
  {Preskill}},\ }\bibfield  {title} {\bibinfo {title} {Quantum computing and
  the entanglement frontier},\ }\href@noop {} {\bibfield  {journal} {\bibinfo
  {journal} {arXiv preprint arXiv:1203.5813}\ } (\bibinfo {year}
  {2012})}\BibitemShut {NoStop}%
\bibitem [{\citenamefont {Preskill}(2018)}]{preskill2018quantum}%
  \BibitemOpen
  \bibfield  {author} {\bibinfo {author} {\bibfnamefont {J.}~\bibnamefont
  {Preskill}},\ }\bibfield  {title} {\bibinfo {title} {Quantum computing in the
  nisq era and beyond},\ }\href@noop {} {\bibfield  {journal} {\bibinfo
  {journal} {Quantum}\ }\textbf {\bibinfo {volume} {2}},\ \bibinfo {pages} {79}
  (\bibinfo {year} {2018})}\BibitemShut {NoStop}%
\bibitem [{\citenamefont {Boixo}\ \emph {et~al.}(2018)\citenamefont {Boixo},
  \citenamefont {Isakov}, \citenamefont {Smelyanskiy}, \citenamefont {Babbush},
  \citenamefont {Ding}, \citenamefont {Jiang}, \citenamefont {Bremner},
  \citenamefont {Martinis},\ and\ \citenamefont
  {Neven}}]{boixo2018characterizing}%
  \BibitemOpen
  \bibfield  {author} {\bibinfo {author} {\bibfnamefont {S.}~\bibnamefont
  {Boixo}}, \bibinfo {author} {\bibfnamefont {S.~V.}\ \bibnamefont {Isakov}},
  \bibinfo {author} {\bibfnamefont {V.~N.}\ \bibnamefont {Smelyanskiy}},
  \bibinfo {author} {\bibfnamefont {R.}~\bibnamefont {Babbush}}, \bibinfo
  {author} {\bibfnamefont {N.}~\bibnamefont {Ding}}, \bibinfo {author}
  {\bibfnamefont {Z.}~\bibnamefont {Jiang}}, \bibinfo {author} {\bibfnamefont
  {M.~J.}\ \bibnamefont {Bremner}}, \bibinfo {author} {\bibfnamefont {J.~M.}\
  \bibnamefont {Martinis}},\ and\ \bibinfo {author} {\bibfnamefont
  {H.}~\bibnamefont {Neven}},\ }\bibfield  {title} {\bibinfo {title}
  {Characterizing quantum supremacy in near-term devices},\ }\href@noop {}
  {\bibfield  {journal} {\bibinfo  {journal} {Nature Physics}\ }\textbf
  {\bibinfo {volume} {14}},\ \bibinfo {pages} {595} (\bibinfo {year}
  {2018})}\BibitemShut {NoStop}%
\bibitem [{\citenamefont {Madsen}\ \emph {et~al.}(2022)\citenamefont {Madsen},
  \citenamefont {Laudenbach}, \citenamefont {Askarani}, \citenamefont
  {Rortais}, \citenamefont {Vincent}, \citenamefont {Bulmer}, \citenamefont
  {Miatto}, \citenamefont {Neuhaus}, \citenamefont {Helt}, \citenamefont
  {Collins} \emph {et~al.}}]{madsen2022quantum}%
  \BibitemOpen
  \bibfield  {author} {\bibinfo {author} {\bibfnamefont {L.~S.}\ \bibnamefont
  {Madsen}}, \bibinfo {author} {\bibfnamefont {F.}~\bibnamefont {Laudenbach}},
  \bibinfo {author} {\bibfnamefont {M.~F.}\ \bibnamefont {Askarani}}, \bibinfo
  {author} {\bibfnamefont {F.}~\bibnamefont {Rortais}}, \bibinfo {author}
  {\bibfnamefont {T.}~\bibnamefont {Vincent}}, \bibinfo {author} {\bibfnamefont
  {J.~F.}\ \bibnamefont {Bulmer}}, \bibinfo {author} {\bibfnamefont {F.~M.}\
  \bibnamefont {Miatto}}, \bibinfo {author} {\bibfnamefont {L.}~\bibnamefont
  {Neuhaus}}, \bibinfo {author} {\bibfnamefont {L.~G.}\ \bibnamefont {Helt}},
  \bibinfo {author} {\bibfnamefont {M.~J.}\ \bibnamefont {Collins}}, \emph
  {et~al.},\ }\bibfield  {title} {\bibinfo {title} {Quantum computational
  advantage with a programmable photonic processor},\ }\href@noop {} {\bibfield
   {journal} {\bibinfo  {journal} {Nature}\ }\textbf {\bibinfo {volume}
  {606}},\ \bibinfo {pages} {75} (\bibinfo {year} {2022})}\BibitemShut
  {NoStop}%
\bibitem [{\citenamefont {Morvan}\ \emph {et~al.}(2023)\citenamefont {Morvan},
  \citenamefont {Villalonga}, \citenamefont {Mi}, \citenamefont {Mandr{\`a}},
  \citenamefont {Bengtsson}, \citenamefont {Klimov}, \citenamefont {Chen},
  \citenamefont {Hong}, \citenamefont {Erickson}, \citenamefont {Drozdov} \emph
  {et~al.}}]{morvan2023phase}%
  \BibitemOpen
  \bibfield  {author} {\bibinfo {author} {\bibfnamefont {A.}~\bibnamefont
  {Morvan}}, \bibinfo {author} {\bibfnamefont {B.}~\bibnamefont {Villalonga}},
  \bibinfo {author} {\bibfnamefont {X.}~\bibnamefont {Mi}}, \bibinfo {author}
  {\bibfnamefont {S.}~\bibnamefont {Mandr{\`a}}}, \bibinfo {author}
  {\bibfnamefont {A.}~\bibnamefont {Bengtsson}}, \bibinfo {author}
  {\bibfnamefont {P.}~\bibnamefont {Klimov}}, \bibinfo {author} {\bibfnamefont
  {Z.}~\bibnamefont {Chen}}, \bibinfo {author} {\bibfnamefont {S.}~\bibnamefont
  {Hong}}, \bibinfo {author} {\bibfnamefont {C.}~\bibnamefont {Erickson}},
  \bibinfo {author} {\bibfnamefont {I.}~\bibnamefont {Drozdov}}, \emph
  {et~al.},\ }\bibfield  {title} {\bibinfo {title} {Phase transition in random
  circuit sampling},\ }\href@noop {} {\bibfield  {journal} {\bibinfo  {journal}
  {arXiv preprint arXiv:2304.11119}\ } (\bibinfo {year} {2023})}\BibitemShut
  {NoStop}%
\bibitem [{\citenamefont {Maring}\ \emph {et~al.}(2023)\citenamefont {Maring},
  \citenamefont {Fyrillas}, \citenamefont {Pont}, \citenamefont {Ivanov},
  \citenamefont {Stepanov}, \citenamefont {Margaria}, \citenamefont {Hease},
  \citenamefont {Pishchagin}, \citenamefont {Au}, \citenamefont {Boissier}
  \emph {et~al.}}]{maring2023generalpurpose}%
  \BibitemOpen
  \bibfield  {author} {\bibinfo {author} {\bibfnamefont {N.}~\bibnamefont
  {Maring}}, \bibinfo {author} {\bibfnamefont {A.}~\bibnamefont {Fyrillas}},
  \bibinfo {author} {\bibfnamefont {M.}~\bibnamefont {Pont}}, \bibinfo {author}
  {\bibfnamefont {E.}~\bibnamefont {Ivanov}}, \bibinfo {author} {\bibfnamefont
  {P.}~\bibnamefont {Stepanov}}, \bibinfo {author} {\bibfnamefont
  {N.}~\bibnamefont {Margaria}}, \bibinfo {author} {\bibfnamefont
  {W.}~\bibnamefont {Hease}}, \bibinfo {author} {\bibfnamefont
  {A.}~\bibnamefont {Pishchagin}}, \bibinfo {author} {\bibfnamefont {T.~H.}\
  \bibnamefont {Au}}, \bibinfo {author} {\bibfnamefont {S.}~\bibnamefont
  {Boissier}}, \emph {et~al.},\ }\href
  {https://doi.org/https://doi.org/10.48550/arXiv.2306.00874} {\bibinfo {title}
  {A general-purpose single-photon-based quantum computing platform}} (\bibinfo
  {year} {2023}),\ \Eprint {https://arxiv.org/abs/2306.00874} {arXiv:2306.00874
  [quant-ph]} \BibitemShut {NoStop}%
\bibitem [{\citenamefont {Quantum}\ \emph {et~al.}(2020)\citenamefont
  {Quantum}, \citenamefont {Collaborators*†}, \citenamefont {Arute},
  \citenamefont {Arya}, \citenamefont {Babbush}, \citenamefont {Bacon},
  \citenamefont {Bardin}, \citenamefont {Barends}, \citenamefont {Boixo},
  \citenamefont {Broughton}, \citenamefont {Buckley} \emph
  {et~al.}}]{google2020hartree}%
  \BibitemOpen
  \bibfield  {author} {\bibinfo {author} {\bibfnamefont {G.~A.}\ \bibnamefont
  {Quantum}}, \bibinfo {author} {\bibnamefont {Collaborators*†}}, \bibinfo
  {author} {\bibfnamefont {F.}~\bibnamefont {Arute}}, \bibinfo {author}
  {\bibfnamefont {K.}~\bibnamefont {Arya}}, \bibinfo {author} {\bibfnamefont
  {R.}~\bibnamefont {Babbush}}, \bibinfo {author} {\bibfnamefont
  {D.}~\bibnamefont {Bacon}}, \bibinfo {author} {\bibfnamefont {J.~C.}\
  \bibnamefont {Bardin}}, \bibinfo {author} {\bibfnamefont {R.}~\bibnamefont
  {Barends}}, \bibinfo {author} {\bibfnamefont {S.}~\bibnamefont {Boixo}},
  \bibinfo {author} {\bibfnamefont {M.}~\bibnamefont {Broughton}}, \bibinfo
  {author} {\bibfnamefont {B.~B.}\ \bibnamefont {Buckley}}, \emph {et~al.},\
  }\bibfield  {title} {\bibinfo {title} {Hartree-fock on a superconducting
  qubit quantum computer},\ }\href {https://doi.org/DOI:
  10.1126/science.abb9811} {\bibfield  {journal} {\bibinfo  {journal}
  {Science}\ }\textbf {\bibinfo {volume} {369}},\ \bibinfo {pages} {1084}
  (\bibinfo {year} {2020})}\BibitemShut {NoStop}%
\bibitem [{\citenamefont {Aaronson}\ and\ \citenamefont
  {Arkhipov}(2011)}]{aaronson2011computational}%
  \BibitemOpen
  \bibfield  {author} {\bibinfo {author} {\bibfnamefont {S.}~\bibnamefont
  {Aaronson}}\ and\ \bibinfo {author} {\bibfnamefont {A.}~\bibnamefont
  {Arkhipov}},\ }\bibfield  {title} {\bibinfo {title} {The computational
  complexity of linear optics},\ }in\ \href@noop {} {\emph {\bibinfo
  {booktitle} {Proceedings of the forty-third annual ACM symposium on Theory of
  computing}}}\ (\bibinfo {year} {2011})\ pp.\ \bibinfo {pages}
  {333--342}\BibitemShut {NoStop}%
\bibitem [{\citenamefont {Hamilton}\ \emph {et~al.}(2017)\citenamefont
  {Hamilton}, \citenamefont {Kruse}, \citenamefont {Sansoni}, \citenamefont
  {Barkhofen}, \citenamefont {Silberhorn},\ and\ \citenamefont
  {Jex}}]{hamilton2017gaussian}%
  \BibitemOpen
  \bibfield  {author} {\bibinfo {author} {\bibfnamefont {C.~S.}\ \bibnamefont
  {Hamilton}}, \bibinfo {author} {\bibfnamefont {R.}~\bibnamefont {Kruse}},
  \bibinfo {author} {\bibfnamefont {L.}~\bibnamefont {Sansoni}}, \bibinfo
  {author} {\bibfnamefont {S.}~\bibnamefont {Barkhofen}}, \bibinfo {author}
  {\bibfnamefont {C.}~\bibnamefont {Silberhorn}},\ and\ \bibinfo {author}
  {\bibfnamefont {I.}~\bibnamefont {Jex}},\ }\bibfield  {title} {\bibinfo
  {title} {Gaussian boson sampling},\ }\href@noop {} {\bibfield  {journal}
  {\bibinfo  {journal} {Physical review letters}\ }\textbf {\bibinfo {volume}
  {119}},\ \bibinfo {pages} {170501} (\bibinfo {year} {2017})}\BibitemShut
  {NoStop}%
\bibitem [{\citenamefont {Bartolucci}\ \emph {et~al.}(2023)\citenamefont
  {Bartolucci}, \citenamefont {Birchall}, \citenamefont {Bombin}, \citenamefont
  {Cable}, \citenamefont {Dawson}, \citenamefont {Gimeno-Segovia},
  \citenamefont {Johnston}, \citenamefont {Kieling}, \citenamefont {Nickerson},
  \citenamefont {Pant} \emph {et~al.}}]{bartolucci2023fusion}%
  \BibitemOpen
  \bibfield  {author} {\bibinfo {author} {\bibfnamefont {S.}~\bibnamefont
  {Bartolucci}}, \bibinfo {author} {\bibfnamefont {P.}~\bibnamefont
  {Birchall}}, \bibinfo {author} {\bibfnamefont {H.}~\bibnamefont {Bombin}},
  \bibinfo {author} {\bibfnamefont {H.}~\bibnamefont {Cable}}, \bibinfo
  {author} {\bibfnamefont {C.}~\bibnamefont {Dawson}}, \bibinfo {author}
  {\bibfnamefont {M.}~\bibnamefont {Gimeno-Segovia}}, \bibinfo {author}
  {\bibfnamefont {E.}~\bibnamefont {Johnston}}, \bibinfo {author}
  {\bibfnamefont {K.}~\bibnamefont {Kieling}}, \bibinfo {author} {\bibfnamefont
  {N.}~\bibnamefont {Nickerson}}, \bibinfo {author} {\bibfnamefont
  {M.}~\bibnamefont {Pant}}, \emph {et~al.},\ }\bibfield  {title} {\bibinfo
  {title} {Fusion-based quantum computation},\ }\href
  {https://doi.org/https://doi.org/10.1038/s41467-023-36493-1} {\bibfield
  {journal} {\bibinfo  {journal} {Nature Communications}\ }\textbf {\bibinfo
  {volume} {14}},\ \bibinfo {pages} {912} (\bibinfo {year} {2023})}\BibitemShut
  {NoStop}%
\bibitem [{\citenamefont {Paesani}\ and\ \citenamefont
  {Brown}(2023)}]{Paesani2023}%
  \BibitemOpen
  \bibfield  {author} {\bibinfo {author} {\bibfnamefont {S.}~\bibnamefont
  {Paesani}}\ and\ \bibinfo {author} {\bibfnamefont {B.~J.}\ \bibnamefont
  {Brown}},\ }\bibfield  {title} {\bibinfo {title} {High-threshold quantum
  computing by fusing one-dimensional cluster states},\ }\href
  {https://doi.org/10.1103/PhysRevLett.131.120603} {\bibfield  {journal}
  {\bibinfo  {journal} {Phys. Rev. Lett.}\ }\textbf {\bibinfo {volume} {131}},\
  \bibinfo {pages} {120603} (\bibinfo {year} {2023})}\BibitemShut {NoStop}%
\bibitem [{\citenamefont {Valiant}(1979)}]{valiant1979complexity}%
  \BibitemOpen
  \bibfield  {author} {\bibinfo {author} {\bibfnamefont {L.~G.}\ \bibnamefont
  {Valiant}},\ }\bibfield  {title} {\bibinfo {title} {The complexity of
  computing the permanent},\ }\href@noop {} {\bibfield  {journal} {\bibinfo
  {journal} {Theoretical computer science}\ }\textbf {\bibinfo {volume} {8}},\
  \bibinfo {pages} {189} (\bibinfo {year} {1979})}\BibitemShut {NoStop}%
\bibitem [{\citenamefont {Scheel}(2004)}]{Scheel2004}%
  \BibitemOpen
  \bibfield  {author} {\bibinfo {author} {\bibfnamefont {S.}~\bibnamefont
  {Scheel}},\ }\bibfield  {title} {\bibinfo {title} {Permanents in linear
  optical networks},\ }\href@noop {} {\bibfield  {journal} {\bibinfo  {journal}
  {arXiv preprint arXiv:quantph/0406127}\ } (\bibinfo {year}
  {2004})}\BibitemShut {NoStop}%
\bibitem [{\citenamefont {Brod}\ \emph {et~al.}(2019)\citenamefont {Brod},
  \citenamefont {Galv{\~a}o}, \citenamefont {Crespi}, \citenamefont {Osellame},
  \citenamefont {Spagnolo},\ and\ \citenamefont
  {Sciarrino}}]{brod2019photonic}%
  \BibitemOpen
  \bibfield  {author} {\bibinfo {author} {\bibfnamefont {D.~J.}\ \bibnamefont
  {Brod}}, \bibinfo {author} {\bibfnamefont {E.~F.}\ \bibnamefont
  {Galv{\~a}o}}, \bibinfo {author} {\bibfnamefont {A.}~\bibnamefont {Crespi}},
  \bibinfo {author} {\bibfnamefont {R.}~\bibnamefont {Osellame}}, \bibinfo
  {author} {\bibfnamefont {N.}~\bibnamefont {Spagnolo}},\ and\ \bibinfo
  {author} {\bibfnamefont {F.}~\bibnamefont {Sciarrino}},\ }\bibfield  {title}
  {\bibinfo {title} {Photonic implementation of boson sampling: a review},\
  }\href@noop {} {\bibfield  {journal} {\bibinfo  {journal} {Advanced
  Photonics}\ }\textbf {\bibinfo {volume} {1}},\ \bibinfo {pages} {034001}
  (\bibinfo {year} {2019})}\BibitemShut {NoStop}%
\bibitem [{\citenamefont {Renema}\ \emph
  {et~al.}(2018{\natexlab{a}})\citenamefont {Renema}, \citenamefont
  {Shchesnovich},\ and\ \citenamefont {Garcia-Patron}}]{renema2018classical}%
  \BibitemOpen
  \bibfield  {author} {\bibinfo {author} {\bibfnamefont {J.}~\bibnamefont
  {Renema}}, \bibinfo {author} {\bibfnamefont {V.}~\bibnamefont
  {Shchesnovich}},\ and\ \bibinfo {author} {\bibfnamefont {R.}~\bibnamefont
  {Garcia-Patron}},\ }\bibfield  {title} {\bibinfo {title} {Classical
  simulability of noisy boson sampling},\ }\href@noop {} {\bibfield  {journal}
  {\bibinfo  {journal} {arXiv preprint arXiv:1809.01953}\ } (\bibinfo {year}
  {2018}{\natexlab{a}})}\BibitemShut {NoStop}%
\bibitem [{\citenamefont {Renema}\ \emph
  {et~al.}(2018{\natexlab{b}})\citenamefont {Renema}, \citenamefont {Menssen},
  \citenamefont {Clements}, \citenamefont {Triginer}, \citenamefont
  {Kolthammer},\ and\ \citenamefont {Walmsley}}]{renema2018efficient}%
  \BibitemOpen
  \bibfield  {author} {\bibinfo {author} {\bibfnamefont {J.~J.}\ \bibnamefont
  {Renema}}, \bibinfo {author} {\bibfnamefont {A.}~\bibnamefont {Menssen}},
  \bibinfo {author} {\bibfnamefont {W.~R.}\ \bibnamefont {Clements}}, \bibinfo
  {author} {\bibfnamefont {G.}~\bibnamefont {Triginer}}, \bibinfo {author}
  {\bibfnamefont {W.~S.}\ \bibnamefont {Kolthammer}},\ and\ \bibinfo {author}
  {\bibfnamefont {I.~A.}\ \bibnamefont {Walmsley}},\ }\bibfield  {title}
  {\bibinfo {title} {Efficient classical algorithm for boson sampling with
  partially distinguishable photons},\ }\href@noop {} {\bibfield  {journal}
  {\bibinfo  {journal} {Physical review letters}\ }\textbf {\bibinfo {volume}
  {120}},\ \bibinfo {pages} {220502} (\bibinfo {year}
  {2018}{\natexlab{b}})}\BibitemShut {NoStop}%
\bibitem [{\citenamefont {Elshaari}\ \emph {et~al.}(2020)\citenamefont
  {Elshaari}, \citenamefont {Pernice}, \citenamefont {Srinivasan},
  \citenamefont {Benson},\ and\ \citenamefont {Zwiller}}]{Elshaari2020np}%
  \BibitemOpen
  \bibfield  {author} {\bibinfo {author} {\bibfnamefont {A.~W.}\ \bibnamefont
  {Elshaari}}, \bibinfo {author} {\bibfnamefont {W.}~\bibnamefont {Pernice}},
  \bibinfo {author} {\bibfnamefont {K.}~\bibnamefont {Srinivasan}}, \bibinfo
  {author} {\bibfnamefont {O.}~\bibnamefont {Benson}},\ and\ \bibinfo {author}
  {\bibfnamefont {V.}~\bibnamefont {Zwiller}},\ }\bibfield  {title} {\bibinfo
  {title} {Hybrid integrated quantum photonic circuits},\ }\href@noop {}
  {\bibfield  {journal} {\bibinfo  {journal} {Nat. Photon.}\ }\textbf {\bibinfo
  {volume} {14}},\ \bibinfo {pages} {285} (\bibinfo {year} {2020})}\BibitemShut
  {NoStop}%
\bibitem [{\citenamefont {Uppu}\ \emph {et~al.}(2021)\citenamefont {Uppu},
  \citenamefont {Midolo}, \citenamefont {Zhou}, \citenamefont {Carolan},\ and\
  \citenamefont {Lodahl}}]{uppu2021natnano}%
  \BibitemOpen
  \bibfield  {author} {\bibinfo {author} {\bibfnamefont {R.}~\bibnamefont
  {Uppu}}, \bibinfo {author} {\bibfnamefont {L.}~\bibnamefont {Midolo}},
  \bibinfo {author} {\bibfnamefont {X.}~\bibnamefont {Zhou}}, \bibinfo {author}
  {\bibfnamefont {J.}~\bibnamefont {Carolan}},\ and\ \bibinfo {author}
  {\bibfnamefont {P.}~\bibnamefont {Lodahl}},\ }\bibfield  {title} {\bibinfo
  {title} {Quantum-dot-based deterministic photon–emitter interfaces for
  scalable photonic quantum technology},\ }\href
  {https://doi.org/https://dx.doi.org/10.1038/s41565-021-00965-6} {\bibfield
  {journal} {\bibinfo  {journal} {Nat. Nanotechnol.}\ }\textbf {\bibinfo
  {volume} {16}},\ \bibinfo {pages} {1308} (\bibinfo {year}
  {2021})}\BibitemShut {NoStop}%
\bibitem [{\citenamefont {Pelucchi}\ \emph {et~al.}(2021)\citenamefont
  {Pelucchi}, \citenamefont {Fagas}, \citenamefont {Aharonovich}, \citenamefont
  {Englund}, \citenamefont {Figueroa}, \citenamefont {Gong}, \citenamefont
  {Hannes}, \citenamefont {Liu}, \citenamefont {Lu}, \citenamefont {Matsuda}
  \emph {et~al.}}]{Pelucchi2021}%
  \BibitemOpen
  \bibfield  {author} {\bibinfo {author} {\bibfnamefont {E.}~\bibnamefont
  {Pelucchi}}, \bibinfo {author} {\bibfnamefont {G.}~\bibnamefont {Fagas}},
  \bibinfo {author} {\bibfnamefont {I.}~\bibnamefont {Aharonovich}}, \bibinfo
  {author} {\bibfnamefont {D.}~\bibnamefont {Englund}}, \bibinfo {author}
  {\bibfnamefont {E.}~\bibnamefont {Figueroa}}, \bibinfo {author}
  {\bibfnamefont {Q.}~\bibnamefont {Gong}}, \bibinfo {author} {\bibfnamefont
  {H.}~\bibnamefont {Hannes}}, \bibinfo {author} {\bibfnamefont
  {J.}~\bibnamefont {Liu}}, \bibinfo {author} {\bibfnamefont {C.-Y.}\
  \bibnamefont {Lu}}, \bibinfo {author} {\bibfnamefont {N.}~\bibnamefont
  {Matsuda}}, \emph {et~al.},\ }\bibfield  {title} {\bibinfo {title} {The
  potential and global outlook of integrated photonics for quantum
  technologies},\ }\href {https://doi.org/10.1038/s42254-021-00398-z}
  {\bibfield  {journal} {\bibinfo  {journal} {Nat. Rev. Phys.}\ }\textbf
  {\bibinfo {volume} {4}},\ \bibinfo {pages} {194} (\bibinfo {year}
  {2021})}\BibitemShut {NoStop}%
\bibitem [{\citenamefont {Moody}\ \emph {et~al.}(2022)\citenamefont {Moody},
  \citenamefont {Sorger}, \citenamefont {Blumenthal}, \citenamefont
  {Juodawlkis}, \citenamefont {Loh}, \citenamefont {Sorace-Agaskar},
  \citenamefont {Jones}, \citenamefont {Balram}, \citenamefont {Matthews},
  \citenamefont {Laing} \emph {et~al.}}]{Moody2022}%
  \BibitemOpen
  \bibfield  {author} {\bibinfo {author} {\bibfnamefont {G.}~\bibnamefont
  {Moody}}, \bibinfo {author} {\bibfnamefont {V.~J.}\ \bibnamefont {Sorger}},
  \bibinfo {author} {\bibfnamefont {D.~J.}\ \bibnamefont {Blumenthal}},
  \bibinfo {author} {\bibfnamefont {P.~W.}\ \bibnamefont {Juodawlkis}},
  \bibinfo {author} {\bibfnamefont {W.}~\bibnamefont {Loh}}, \bibinfo {author}
  {\bibfnamefont {C.}~\bibnamefont {Sorace-Agaskar}}, \bibinfo {author}
  {\bibfnamefont {A.~E.}\ \bibnamefont {Jones}}, \bibinfo {author}
  {\bibfnamefont {K.~C.}\ \bibnamefont {Balram}}, \bibinfo {author}
  {\bibfnamefont {J.~C.~F.}\ \bibnamefont {Matthews}}, \bibinfo {author}
  {\bibfnamefont {A.}~\bibnamefont {Laing}}, \emph {et~al.},\ }\bibfield
  {title} {\bibinfo {title} {{2022 Roadmap on integrated quantum photonics}},\
  }\href {https://doi.org/10.1088/2515-7647/ac1ef4} {\bibfield  {journal}
  {\bibinfo  {journal} {J. Phys. Photonics}\ }\textbf {\bibinfo {volume} {4}},\
  \bibinfo {pages} {012501} (\bibinfo {year} {2022})}\BibitemShut {NoStop}%
\bibitem [{\citenamefont {Broome}\ \emph {et~al.}(2013)\citenamefont {Broome},
  \citenamefont {Fedrizzi}, \citenamefont {Rahimi-Keshari}, \citenamefont
  {Dove}, \citenamefont {Aaronson}, \citenamefont {Ralph},\ and\ \citenamefont
  {White}}]{broome2013photonic}%
  \BibitemOpen
  \bibfield  {author} {\bibinfo {author} {\bibfnamefont {M.~A.}\ \bibnamefont
  {Broome}}, \bibinfo {author} {\bibfnamefont {A.}~\bibnamefont {Fedrizzi}},
  \bibinfo {author} {\bibfnamefont {S.}~\bibnamefont {Rahimi-Keshari}},
  \bibinfo {author} {\bibfnamefont {J.}~\bibnamefont {Dove}}, \bibinfo {author}
  {\bibfnamefont {S.}~\bibnamefont {Aaronson}}, \bibinfo {author}
  {\bibfnamefont {T.~C.}\ \bibnamefont {Ralph}},\ and\ \bibinfo {author}
  {\bibfnamefont {A.~G.}\ \bibnamefont {White}},\ }\bibfield  {title} {\bibinfo
  {title} {Photonic boson sampling in a tunable circuit},\ }\href@noop {}
  {\bibfield  {journal} {\bibinfo  {journal} {Science}\ }\textbf {\bibinfo
  {volume} {339}},\ \bibinfo {pages} {794} (\bibinfo {year}
  {2013})}\BibitemShut {NoStop}%
\bibitem [{\citenamefont {Spring}\ \emph {et~al.}(2013)\citenamefont {Spring},
  \citenamefont {Metcalf}, \citenamefont {Humphreys}, \citenamefont
  {Kolthammer}, \citenamefont {Jin}, \citenamefont {Barbieri}, \citenamefont
  {Datta}, \citenamefont {Thomas-Peter}, \citenamefont {Langford},
  \citenamefont {Kundys} \emph {et~al.}}]{spring2013boson}%
  \BibitemOpen
  \bibfield  {author} {\bibinfo {author} {\bibfnamefont {J.~B.}\ \bibnamefont
  {Spring}}, \bibinfo {author} {\bibfnamefont {B.~J.}\ \bibnamefont {Metcalf}},
  \bibinfo {author} {\bibfnamefont {P.~C.}\ \bibnamefont {Humphreys}}, \bibinfo
  {author} {\bibfnamefont {W.~S.}\ \bibnamefont {Kolthammer}}, \bibinfo
  {author} {\bibfnamefont {X.-M.}\ \bibnamefont {Jin}}, \bibinfo {author}
  {\bibfnamefont {M.}~\bibnamefont {Barbieri}}, \bibinfo {author}
  {\bibfnamefont {A.}~\bibnamefont {Datta}}, \bibinfo {author} {\bibfnamefont
  {N.}~\bibnamefont {Thomas-Peter}}, \bibinfo {author} {\bibfnamefont {N.~K.}\
  \bibnamefont {Langford}}, \bibinfo {author} {\bibfnamefont {D.}~\bibnamefont
  {Kundys}}, \emph {et~al.},\ }\bibfield  {title} {\bibinfo {title} {Boson
  sampling on a photonic chip},\ }\href@noop {} {\bibfield  {journal} {\bibinfo
   {journal} {Science}\ }\textbf {\bibinfo {volume} {339}},\ \bibinfo {pages}
  {798} (\bibinfo {year} {2013})}\BibitemShut {NoStop}%
\bibitem [{\citenamefont {Crespi}\ \emph {et~al.}(2013)\citenamefont {Crespi},
  \citenamefont {Osellame}, \citenamefont {Ramponi}, \citenamefont {Brod},
  \citenamefont {Galvao}, \citenamefont {Spagnolo}, \citenamefont {Vitelli},
  \citenamefont {Maiorino}, \citenamefont {Mataloni},\ and\ \citenamefont
  {Sciarrino}}]{crespi2013integrated}%
  \BibitemOpen
  \bibfield  {author} {\bibinfo {author} {\bibfnamefont {A.}~\bibnamefont
  {Crespi}}, \bibinfo {author} {\bibfnamefont {R.}~\bibnamefont {Osellame}},
  \bibinfo {author} {\bibfnamefont {R.}~\bibnamefont {Ramponi}}, \bibinfo
  {author} {\bibfnamefont {D.~J.}\ \bibnamefont {Brod}}, \bibinfo {author}
  {\bibfnamefont {E.~F.}\ \bibnamefont {Galvao}}, \bibinfo {author}
  {\bibfnamefont {N.}~\bibnamefont {Spagnolo}}, \bibinfo {author}
  {\bibfnamefont {C.}~\bibnamefont {Vitelli}}, \bibinfo {author} {\bibfnamefont
  {E.}~\bibnamefont {Maiorino}}, \bibinfo {author} {\bibfnamefont
  {P.}~\bibnamefont {Mataloni}},\ and\ \bibinfo {author} {\bibfnamefont
  {F.}~\bibnamefont {Sciarrino}},\ }\bibfield  {title} {\bibinfo {title}
  {Integrated multimode interferometers with arbitrary designs for photonic
  boson sampling},\ }\href@noop {} {\bibfield  {journal} {\bibinfo  {journal}
  {Nature photonics}\ }\textbf {\bibinfo {volume} {7}},\ \bibinfo {pages} {545}
  (\bibinfo {year} {2013})}\BibitemShut {NoStop}%
\bibitem [{\citenamefont {Tillmann}\ \emph {et~al.}(2013)\citenamefont
  {Tillmann}, \citenamefont {Daki{\'c}}, \citenamefont {Heilmann},
  \citenamefont {Nolte}, \citenamefont {Szameit},\ and\ \citenamefont
  {Walther}}]{tillmann2013experimental}%
  \BibitemOpen
  \bibfield  {author} {\bibinfo {author} {\bibfnamefont {M.}~\bibnamefont
  {Tillmann}}, \bibinfo {author} {\bibfnamefont {B.}~\bibnamefont {Daki{\'c}}},
  \bibinfo {author} {\bibfnamefont {R.}~\bibnamefont {Heilmann}}, \bibinfo
  {author} {\bibfnamefont {S.}~\bibnamefont {Nolte}}, \bibinfo {author}
  {\bibfnamefont {A.}~\bibnamefont {Szameit}},\ and\ \bibinfo {author}
  {\bibfnamefont {P.}~\bibnamefont {Walther}},\ }\bibfield  {title} {\bibinfo
  {title} {Experimental boson sampling},\ }\href@noop {} {\bibfield  {journal}
  {\bibinfo  {journal} {Nature photonics}\ }\textbf {\bibinfo {volume} {7}},\
  \bibinfo {pages} {540} (\bibinfo {year} {2013})}\BibitemShut {NoStop}%
\bibitem [{\citenamefont {Wang}\ \emph {et~al.}(2019)\citenamefont {Wang},
  \citenamefont {Qin}, \citenamefont {Ding}, \citenamefont {Chen},
  \citenamefont {Chen}, \citenamefont {You}, \citenamefont {He}, \citenamefont
  {Jiang}, \citenamefont {You}, \citenamefont {Wang} \emph
  {et~al.}}]{wang2019boson}%
  \BibitemOpen
  \bibfield  {author} {\bibinfo {author} {\bibfnamefont {H.}~\bibnamefont
  {Wang}}, \bibinfo {author} {\bibfnamefont {J.}~\bibnamefont {Qin}}, \bibinfo
  {author} {\bibfnamefont {X.}~\bibnamefont {Ding}}, \bibinfo {author}
  {\bibfnamefont {M.-C.}\ \bibnamefont {Chen}}, \bibinfo {author}
  {\bibfnamefont {S.}~\bibnamefont {Chen}}, \bibinfo {author} {\bibfnamefont
  {X.}~\bibnamefont {You}}, \bibinfo {author} {\bibfnamefont {Y.-M.}\
  \bibnamefont {He}}, \bibinfo {author} {\bibfnamefont {X.}~\bibnamefont
  {Jiang}}, \bibinfo {author} {\bibfnamefont {L.}~\bibnamefont {You}}, \bibinfo
  {author} {\bibfnamefont {Z.}~\bibnamefont {Wang}}, \emph {et~al.},\
  }\bibfield  {title} {\bibinfo {title} {Boson sampling with 20 input photons
  and a 60-mode interferometer in a $10^{14}$-dimensional hilbert space},\
  }\href {https://doi.org/https://doi.org/10.1103/PhysRevLett.123.250503}
  {\bibfield  {journal} {\bibinfo  {journal} {Physical review letters}\
  }\textbf {\bibinfo {volume} {123}},\ \bibinfo {pages} {250503} (\bibinfo
  {year} {2019})}\BibitemShut {NoStop}%
\bibitem [{\citenamefont {Uppu}\ \emph {et~al.}(2020)\citenamefont {Uppu},
  \citenamefont {Pedersen}, \citenamefont {Wang}, \citenamefont {Olesen},
  \citenamefont {Papon}, \citenamefont {Zhou}, \citenamefont {Midolo},
  \citenamefont {Scholz}, \citenamefont {Wieck}, \citenamefont {Ludwig},\ and\
  \citenamefont {Lodahl}}]{uppu2020scalable}%
  \BibitemOpen
  \bibfield  {author} {\bibinfo {author} {\bibfnamefont {R.}~\bibnamefont
  {Uppu}}, \bibinfo {author} {\bibfnamefont {F.~T.}\ \bibnamefont {Pedersen}},
  \bibinfo {author} {\bibfnamefont {Y.}~\bibnamefont {Wang}}, \bibinfo {author}
  {\bibfnamefont {C.~T.}\ \bibnamefont {Olesen}}, \bibinfo {author}
  {\bibfnamefont {C.}~\bibnamefont {Papon}}, \bibinfo {author} {\bibfnamefont
  {X.}~\bibnamefont {Zhou}}, \bibinfo {author} {\bibfnamefont {L.}~\bibnamefont
  {Midolo}}, \bibinfo {author} {\bibfnamefont {S.}~\bibnamefont {Scholz}},
  \bibinfo {author} {\bibfnamefont {A.~D.}\ \bibnamefont {Wieck}}, \bibinfo
  {author} {\bibfnamefont {A.}~\bibnamefont {Ludwig}},\ and\ \bibinfo {author}
  {\bibfnamefont {P.}~\bibnamefont {Lodahl}},\ }\bibfield  {title} {\bibinfo
  {title} {Scalable integrated single-photon source},\ }\href@noop {}
  {\bibfield  {journal} {\bibinfo  {journal} {Science advances}\ }\textbf
  {\bibinfo {volume} {6}},\ \bibinfo {pages} {eabc8268} (\bibinfo {year}
  {2020})}\BibitemShut {NoStop}%
\bibitem [{\citenamefont {Tomm}\ \emph {et~al.}(2021)\citenamefont {Tomm},
  \citenamefont {Javadi}, \citenamefont {Antoniadis}, \citenamefont {Najer},
  \citenamefont {L{\"o}bl}, \citenamefont {Korsch}, \citenamefont {Schott},
  \citenamefont {Valentin}, \citenamefont {Wieck}, \citenamefont {Ludwig} \emph
  {et~al.}}]{tomm2021nn}%
  \BibitemOpen
  \bibfield  {author} {\bibinfo {author} {\bibfnamefont {N.}~\bibnamefont
  {Tomm}}, \bibinfo {author} {\bibfnamefont {A.}~\bibnamefont {Javadi}},
  \bibinfo {author} {\bibfnamefont {N.~O.}\ \bibnamefont {Antoniadis}},
  \bibinfo {author} {\bibfnamefont {D.}~\bibnamefont {Najer}}, \bibinfo
  {author} {\bibfnamefont {M.~C.}\ \bibnamefont {L{\"o}bl}}, \bibinfo {author}
  {\bibfnamefont {A.~R.}\ \bibnamefont {Korsch}}, \bibinfo {author}
  {\bibfnamefont {R.}~\bibnamefont {Schott}}, \bibinfo {author} {\bibfnamefont
  {S.~R.}\ \bibnamefont {Valentin}}, \bibinfo {author} {\bibfnamefont {A.~D.}\
  \bibnamefont {Wieck}}, \bibinfo {author} {\bibfnamefont {A.}~\bibnamefont
  {Ludwig}}, \emph {et~al.},\ }\bibfield  {title} {\bibinfo {title} {A bright
  and fast source of coherent single photons},\ }\href
  {https://doi.org/https://doi.org/10.1038/s41565-020-00831-x} {\bibfield
  {journal} {\bibinfo  {journal} {Nat. Nanotechnol.}\ }\textbf {\bibinfo
  {volume} {16}},\ \bibinfo {pages} {399} (\bibinfo {year} {2021})}\BibitemShut
  {NoStop}%
\bibitem [{\citenamefont {Stojanovi{\'c}}\ \emph {et~al.}(2018)\citenamefont
  {Stojanovi{\'c}}, \citenamefont {Ram}, \citenamefont {Popovi{\'c}},
  \citenamefont {Lin}, \citenamefont {Moazeni}, \citenamefont {Wade},
  \citenamefont {Sun}, \citenamefont {Alloatti}, \citenamefont {Atabaki},
  \citenamefont {Pavanello} \emph {et~al.}}]{stojanovic2018monolithic}%
  \BibitemOpen
  \bibfield  {author} {\bibinfo {author} {\bibfnamefont {V.}~\bibnamefont
  {Stojanovi{\'c}}}, \bibinfo {author} {\bibfnamefont {R.~J.}\ \bibnamefont
  {Ram}}, \bibinfo {author} {\bibfnamefont {M.}~\bibnamefont {Popovi{\'c}}},
  \bibinfo {author} {\bibfnamefont {S.}~\bibnamefont {Lin}}, \bibinfo {author}
  {\bibfnamefont {S.}~\bibnamefont {Moazeni}}, \bibinfo {author} {\bibfnamefont
  {M.}~\bibnamefont {Wade}}, \bibinfo {author} {\bibfnamefont {C.}~\bibnamefont
  {Sun}}, \bibinfo {author} {\bibfnamefont {L.}~\bibnamefont {Alloatti}},
  \bibinfo {author} {\bibfnamefont {A.}~\bibnamefont {Atabaki}}, \bibinfo
  {author} {\bibfnamefont {F.}~\bibnamefont {Pavanello}}, \emph {et~al.},\
  }\bibfield  {title} {\bibinfo {title} {Monolithic silicon-photonic platforms
  in state-of-the-art cmos soi processes},\ }\href@noop {} {\bibfield
  {journal} {\bibinfo  {journal} {Optics express}\ }\textbf {\bibinfo {volume}
  {26}},\ \bibinfo {pages} {13106} (\bibinfo {year} {2018})}\BibitemShut
  {NoStop}%
\bibitem [{\citenamefont {Bao}\ \emph {et~al.}(2023)\citenamefont {Bao},
  \citenamefont {Fu}, \citenamefont {Pramanik}, \citenamefont {Mao},
  \citenamefont {Chi}, \citenamefont {Cao}, \citenamefont {Zhai}, \citenamefont
  {Mao}, \citenamefont {Dai}, \citenamefont {Chen} \emph
  {et~al.}}]{bao2023very}%
  \BibitemOpen
  \bibfield  {author} {\bibinfo {author} {\bibfnamefont {J.}~\bibnamefont
  {Bao}}, \bibinfo {author} {\bibfnamefont {Z.}~\bibnamefont {Fu}}, \bibinfo
  {author} {\bibfnamefont {T.}~\bibnamefont {Pramanik}}, \bibinfo {author}
  {\bibfnamefont {J.}~\bibnamefont {Mao}}, \bibinfo {author} {\bibfnamefont
  {Y.}~\bibnamefont {Chi}}, \bibinfo {author} {\bibfnamefont {Y.}~\bibnamefont
  {Cao}}, \bibinfo {author} {\bibfnamefont {C.}~\bibnamefont {Zhai}}, \bibinfo
  {author} {\bibfnamefont {Y.}~\bibnamefont {Mao}}, \bibinfo {author}
  {\bibfnamefont {T.}~\bibnamefont {Dai}}, \bibinfo {author} {\bibfnamefont
  {X.}~\bibnamefont {Chen}}, \emph {et~al.},\ }\bibfield  {title} {\bibinfo
  {title} {Very-large-scale integrated quantum graph photonics},\ }\href@noop
  {} {\bibfield  {journal} {\bibinfo  {journal} {Nature Photonics}\ ,\ \bibinfo
  {pages} {1}} (\bibinfo {year} {2023})}\BibitemShut {NoStop}%
\bibitem [{\citenamefont {Zhong}\ \emph {et~al.}(2020)\citenamefont {Zhong},
  \citenamefont {Wang}, \citenamefont {Deng}, \citenamefont {Chen},
  \citenamefont {Peng}, \citenamefont {Luo}, \citenamefont {Qin}, \citenamefont
  {Wu}, \citenamefont {Ding}, \citenamefont {Hu} \emph
  {et~al.}}]{zhong2020quantum}%
  \BibitemOpen
  \bibfield  {author} {\bibinfo {author} {\bibfnamefont {H.-S.}\ \bibnamefont
  {Zhong}}, \bibinfo {author} {\bibfnamefont {H.}~\bibnamefont {Wang}},
  \bibinfo {author} {\bibfnamefont {Y.-H.}\ \bibnamefont {Deng}}, \bibinfo
  {author} {\bibfnamefont {M.-C.}\ \bibnamefont {Chen}}, \bibinfo {author}
  {\bibfnamefont {L.-C.}\ \bibnamefont {Peng}}, \bibinfo {author}
  {\bibfnamefont {Y.-H.}\ \bibnamefont {Luo}}, \bibinfo {author} {\bibfnamefont
  {J.}~\bibnamefont {Qin}}, \bibinfo {author} {\bibfnamefont {D.}~\bibnamefont
  {Wu}}, \bibinfo {author} {\bibfnamefont {X.}~\bibnamefont {Ding}}, \bibinfo
  {author} {\bibfnamefont {Y.}~\bibnamefont {Hu}}, \emph {et~al.},\ }\bibfield
  {title} {\bibinfo {title} {Quantum computational advantage using photons},\
  }\href@noop {} {\bibfield  {journal} {\bibinfo  {journal} {Science}\ }\textbf
  {\bibinfo {volume} {370}},\ \bibinfo {pages} {1460} (\bibinfo {year}
  {2020})}\BibitemShut {NoStop}%
\bibitem [{\citenamefont {Arkhipov}\ and\ \citenamefont
  {Kuperberg}(2012)}]{arkhipov2012bosonic}%
  \BibitemOpen
  \bibfield  {author} {\bibinfo {author} {\bibfnamefont {A.}~\bibnamefont
  {Arkhipov}}\ and\ \bibinfo {author} {\bibfnamefont {G.}~\bibnamefont
  {Kuperberg}},\ }\bibfield  {title} {\bibinfo {title} {The bosonic birthday
  paradox},\ }\href@noop {} {\bibfield  {journal} {\bibinfo  {journal}
  {Geometry \& Topology Monographs}\ }\textbf {\bibinfo {volume} {18}},\
  \bibinfo {pages} {10} (\bibinfo {year} {2012})}\BibitemShut {NoStop}%
\bibitem [{\citenamefont {Bulmer}\ \emph {et~al.}(2022)\citenamefont {Bulmer},
  \citenamefont {Paesani}, \citenamefont {Chadwick},\ and\ \citenamefont
  {Quesada}}]{Bulmer2022}%
  \BibitemOpen
  \bibfield  {author} {\bibinfo {author} {\bibfnamefont {J.~F.~F.}\
  \bibnamefont {Bulmer}}, \bibinfo {author} {\bibfnamefont {S.}~\bibnamefont
  {Paesani}}, \bibinfo {author} {\bibfnamefont {R.~S.}\ \bibnamefont
  {Chadwick}},\ and\ \bibinfo {author} {\bibfnamefont {N.}~\bibnamefont
  {Quesada}},\ }\bibfield  {title} {\bibinfo {title} {Threshold detection
  statistics of bosonic states},\ }\href
  {https://doi.org/10.1103/PhysRevA.106.043712} {\bibfield  {journal} {\bibinfo
   {journal} {Phys. Rev. A}\ }\textbf {\bibinfo {volume} {106}},\ \bibinfo
  {pages} {043712} (\bibinfo {year} {2022})}\BibitemShut {NoStop}%
\bibitem [{\citenamefont {Hong}\ \emph {et~al.}(1987)\citenamefont {Hong},
  \citenamefont {Ou},\ and\ \citenamefont {Mandel}}]{hong1987measurement}%
  \BibitemOpen
  \bibfield  {author} {\bibinfo {author} {\bibfnamefont {C.-K.}\ \bibnamefont
  {Hong}}, \bibinfo {author} {\bibfnamefont {Z.-Y.}\ \bibnamefont {Ou}},\ and\
  \bibinfo {author} {\bibfnamefont {L.}~\bibnamefont {Mandel}},\ }\bibfield
  {title} {\bibinfo {title} {Measurement of subpicosecond time intervals
  between two photons by interference},\ }\href
  {https://doi.org/https://doi.org/10.1103/PhysRevLett.59.2044} {\bibfield
  {journal} {\bibinfo  {journal} {Physical review letters}\ }\textbf {\bibinfo
  {volume} {59}},\ \bibinfo {pages} {2044} (\bibinfo {year}
  {1987})}\BibitemShut {NoStop}%
\bibitem [{\citenamefont {Migdall}\ \emph {et~al.}(2002)\citenamefont
  {Migdall}, \citenamefont {Branning},\ and\ \citenamefont
  {Castelletto}}]{migdall2002tailoring}%
  \BibitemOpen
  \bibfield  {author} {\bibinfo {author} {\bibfnamefont {A.~L.}\ \bibnamefont
  {Migdall}}, \bibinfo {author} {\bibfnamefont {D.}~\bibnamefont {Branning}},\
  and\ \bibinfo {author} {\bibfnamefont {S.}~\bibnamefont {Castelletto}},\
  }\bibfield  {title} {\bibinfo {title} {Tailoring single-photon and
  multiphoton probabilities of a single-photon on-demand source},\ }\href@noop
  {} {\bibfield  {journal} {\bibinfo  {journal} {Physical Review A}\ }\textbf
  {\bibinfo {volume} {66}},\ \bibinfo {pages} {053805} (\bibinfo {year}
  {2002})}\BibitemShut {NoStop}%
\bibitem [{\citenamefont {Bartolucci}\ \emph {et~al.}(2021)\citenamefont
  {Bartolucci}, \citenamefont {Birchall}, \citenamefont {Bonneau},
  \citenamefont {Cable}, \citenamefont {Gimeno-Segovia}, \citenamefont
  {Kieling}, \citenamefont {Nickerson}, \citenamefont {Rudolph},\ and\
  \citenamefont {Sparrow}}]{bartolucci2021switch}%
  \BibitemOpen
  \bibfield  {author} {\bibinfo {author} {\bibfnamefont {S.}~\bibnamefont
  {Bartolucci}}, \bibinfo {author} {\bibfnamefont {P.}~\bibnamefont
  {Birchall}}, \bibinfo {author} {\bibfnamefont {D.}~\bibnamefont {Bonneau}},
  \bibinfo {author} {\bibfnamefont {H.}~\bibnamefont {Cable}}, \bibinfo
  {author} {\bibfnamefont {M.}~\bibnamefont {Gimeno-Segovia}}, \bibinfo
  {author} {\bibfnamefont {K.}~\bibnamefont {Kieling}}, \bibinfo {author}
  {\bibfnamefont {N.}~\bibnamefont {Nickerson}}, \bibinfo {author}
  {\bibfnamefont {T.}~\bibnamefont {Rudolph}},\ and\ \bibinfo {author}
  {\bibfnamefont {C.}~\bibnamefont {Sparrow}},\ }\href@noop {} {\bibinfo
  {title} {Switch networks for photonic fusion-based quantum computing}}
  (\bibinfo {year} {2021}),\ \Eprint {https://arxiv.org/abs/2109.13760}
  {arXiv:2109.13760 [quant-ph]} \BibitemShut {NoStop}%
\bibitem [{\citenamefont {Lodahl}\ \emph {et~al.}(2022)\citenamefont {Lodahl},
  \citenamefont {Ludwig},\ and\ \citenamefont
  {Warburton}}]{lodahl2022deterministic}%
  \BibitemOpen
  \bibfield  {author} {\bibinfo {author} {\bibfnamefont {P.}~\bibnamefont
  {Lodahl}}, \bibinfo {author} {\bibfnamefont {A.}~\bibnamefont {Ludwig}},\
  and\ \bibinfo {author} {\bibfnamefont {R.~J.}\ \bibnamefont {Warburton}},\
  }\bibfield  {title} {\bibinfo {title} {A deterministic source of},\ }\href
  {https://doi.org/https://doi.org/10.1063/PT.3.4962} {\bibfield  {journal}
  {\bibinfo  {journal} {Physics Today}\ }\textbf {\bibinfo {volume} {75}},\
  \bibinfo {pages} {3} (\bibinfo {year} {2022})}\BibitemShut {NoStop}%
\bibitem [{\citenamefont {Motes}\ \emph {et~al.}(2014)\citenamefont {Motes},
  \citenamefont {Gilchrist}, \citenamefont {Dowling},\ and\ \citenamefont
  {Rohde}}]{motes2014scalable}%
  \BibitemOpen
  \bibfield  {author} {\bibinfo {author} {\bibfnamefont {K.~R.}\ \bibnamefont
  {Motes}}, \bibinfo {author} {\bibfnamefont {A.}~\bibnamefont {Gilchrist}},
  \bibinfo {author} {\bibfnamefont {J.~P.}\ \bibnamefont {Dowling}},\ and\
  \bibinfo {author} {\bibfnamefont {P.~P.}\ \bibnamefont {Rohde}},\ }\bibfield
  {title} {\bibinfo {title} {Scalable boson sampling with time-bin encoding
  using a loop-based architecture},\ }\href
  {https://doi.org/https://doi.org/10.1103/PhysRevLett.113.120501} {\bibfield
  {journal} {\bibinfo  {journal} {Physical review letters}\ }\textbf {\bibinfo
  {volume} {113}},\ \bibinfo {pages} {120501} (\bibinfo {year}
  {2014})}\BibitemShut {NoStop}%
\bibitem [{\citenamefont {He}\ \emph {et~al.}(2017)\citenamefont {He},
  \citenamefont {Ding}, \citenamefont {Su}, \citenamefont {Huang},
  \citenamefont {Qin}, \citenamefont {Wang}, \citenamefont {Unsleber},
  \citenamefont {Chen}, \citenamefont {Wang}, \citenamefont {He} \emph
  {et~al.}}]{he2017time}%
  \BibitemOpen
  \bibfield  {author} {\bibinfo {author} {\bibfnamefont {Y.}~\bibnamefont
  {He}}, \bibinfo {author} {\bibfnamefont {X.}~\bibnamefont {Ding}}, \bibinfo
  {author} {\bibfnamefont {Z.-E.}\ \bibnamefont {Su}}, \bibinfo {author}
  {\bibfnamefont {H.-L.}\ \bibnamefont {Huang}}, \bibinfo {author}
  {\bibfnamefont {J.}~\bibnamefont {Qin}}, \bibinfo {author} {\bibfnamefont
  {C.}~\bibnamefont {Wang}}, \bibinfo {author} {\bibfnamefont {S.}~\bibnamefont
  {Unsleber}}, \bibinfo {author} {\bibfnamefont {C.}~\bibnamefont {Chen}},
  \bibinfo {author} {\bibfnamefont {H.}~\bibnamefont {Wang}}, \bibinfo {author}
  {\bibfnamefont {Y.-M.}\ \bibnamefont {He}}, \emph {et~al.},\ }\bibfield
  {title} {\bibinfo {title} {Time-bin-encoded boson sampling with a
  single-photon device},\ }\href
  {https://doi.org/https://doi.org/10.1103/PhysRevLett.118.190501} {\bibfield
  {journal} {\bibinfo  {journal} {Physical review letters}\ }\textbf {\bibinfo
  {volume} {118}},\ \bibinfo {pages} {190501} (\bibinfo {year}
  {2017})}\BibitemShut {NoStop}%
\bibitem [{\citenamefont {Hummel}\ \emph {et~al.}(2019)\citenamefont {Hummel},
  \citenamefont {Ouellet-Plamondon}, \citenamefont {Ugur}, \citenamefont
  {Kulkova}, \citenamefont {Lund-Hansen}, \citenamefont {Broome}, \citenamefont
  {Uppu},\ and\ \citenamefont {Lodahl}}]{hummel2019efficient}%
  \BibitemOpen
  \bibfield  {author} {\bibinfo {author} {\bibfnamefont {T.}~\bibnamefont
  {Hummel}}, \bibinfo {author} {\bibfnamefont {C.}~\bibnamefont
  {Ouellet-Plamondon}}, \bibinfo {author} {\bibfnamefont {E.}~\bibnamefont
  {Ugur}}, \bibinfo {author} {\bibfnamefont {I.}~\bibnamefont {Kulkova}},
  \bibinfo {author} {\bibfnamefont {T.}~\bibnamefont {Lund-Hansen}}, \bibinfo
  {author} {\bibfnamefont {M.~A.}\ \bibnamefont {Broome}}, \bibinfo {author}
  {\bibfnamefont {R.}~\bibnamefont {Uppu}},\ and\ \bibinfo {author}
  {\bibfnamefont {P.}~\bibnamefont {Lodahl}},\ }\bibfield  {title} {\bibinfo
  {title} {Efficient demultiplexed single-photon source with a quantum dot
  coupled to a nanophotonic waveguide},\ }\href@noop {} {\bibfield  {journal}
  {\bibinfo  {journal} {Applied Physics Letters}\ }\textbf {\bibinfo {volume}
  {115}},\ \bibinfo {pages} {021102} (\bibinfo {year} {2019})}\BibitemShut
  {NoStop}%
\bibitem [{\citenamefont {Zhai}\ \emph {et~al.}(2022)\citenamefont {Zhai},
  \citenamefont {Nguyen}, \citenamefont {Spinnler}, \citenamefont {Ritzmann},
  \citenamefont {L{\"o}bl}, \citenamefont {Wieck}, \citenamefont {Ludwig},
  \citenamefont {Javadi},\ and\ \citenamefont {Warburton}}]{zhai2022quantum}%
  \BibitemOpen
  \bibfield  {author} {\bibinfo {author} {\bibfnamefont {L.}~\bibnamefont
  {Zhai}}, \bibinfo {author} {\bibfnamefont {G.~N.}\ \bibnamefont {Nguyen}},
  \bibinfo {author} {\bibfnamefont {C.}~\bibnamefont {Spinnler}}, \bibinfo
  {author} {\bibfnamefont {J.}~\bibnamefont {Ritzmann}}, \bibinfo {author}
  {\bibfnamefont {M.~C.}\ \bibnamefont {L{\"o}bl}}, \bibinfo {author}
  {\bibfnamefont {A.~D.}\ \bibnamefont {Wieck}}, \bibinfo {author}
  {\bibfnamefont {A.}~\bibnamefont {Ludwig}}, \bibinfo {author} {\bibfnamefont
  {A.}~\bibnamefont {Javadi}},\ and\ \bibinfo {author} {\bibfnamefont {R.~J.}\
  \bibnamefont {Warburton}},\ }\bibfield  {title} {\bibinfo {title} {Quantum
  interference of identical photons from remote gaas quantum dots},\
  }\href@noop {} {\bibfield  {journal} {\bibinfo  {journal} {Nature
  Nanotechnology}\ }\textbf {\bibinfo {volume} {17}},\ \bibinfo {pages} {829}
  (\bibinfo {year} {2022})}\BibitemShut {NoStop}%
\bibitem [{\citenamefont {Papon}\ \emph {et~al.}(2022)\citenamefont {Papon},
  \citenamefont {Wang}, \citenamefont {Uppu}, \citenamefont {Scholz},
  \citenamefont {Wieck}, \citenamefont {Ludwig}, \citenamefont {Lodahl},\ and\
  \citenamefont {Midolo}}]{papon2022independent}%
  \BibitemOpen
  \bibfield  {author} {\bibinfo {author} {\bibfnamefont {C.}~\bibnamefont
  {Papon}}, \bibinfo {author} {\bibfnamefont {Y.}~\bibnamefont {Wang}},
  \bibinfo {author} {\bibfnamefont {R.}~\bibnamefont {Uppu}}, \bibinfo {author}
  {\bibfnamefont {S.}~\bibnamefont {Scholz}}, \bibinfo {author} {\bibfnamefont
  {A.~D.}\ \bibnamefont {Wieck}}, \bibinfo {author} {\bibfnamefont
  {A.}~\bibnamefont {Ludwig}}, \bibinfo {author} {\bibfnamefont
  {P.}~\bibnamefont {Lodahl}},\ and\ \bibinfo {author} {\bibfnamefont
  {L.}~\bibnamefont {Midolo}},\ }\href
  {https://doi.org/https://doi.org/10.48550/arXiv.2210.09826} {\bibinfo {title}
  {Independent operation of two waveguide-integrated single-photon sources}}
  (\bibinfo {year} {2022}),\ \Eprint {https://arxiv.org/abs/2210.09826}
  {arXiv:2210.09826 [quant-ph]} \BibitemShut {NoStop}%
\bibitem [{\citenamefont {Aaronson}\ and\ \citenamefont
  {Brod}(2016)}]{aaronson2016bosonsampling}%
  \BibitemOpen
  \bibfield  {author} {\bibinfo {author} {\bibfnamefont {S.}~\bibnamefont
  {Aaronson}}\ and\ \bibinfo {author} {\bibfnamefont {D.~J.}\ \bibnamefont
  {Brod}},\ }\bibfield  {title} {\bibinfo {title} {Bosonsampling with lost
  photons},\ }\href@noop {} {\bibfield  {journal} {\bibinfo  {journal}
  {Physical Review A}\ }\textbf {\bibinfo {volume} {93}},\ \bibinfo {pages}
  {012335} (\bibinfo {year} {2016})}\BibitemShut {NoStop}%
\bibitem [{\citenamefont {Carolan}\ \emph {et~al.}(2015)\citenamefont
  {Carolan}, \citenamefont {Harrold}, \citenamefont {Sparrow}, \citenamefont
  {Mart{\'\i}n-L{\'o}pez}, \citenamefont {Russell}, \citenamefont
  {Silverstone}, \citenamefont {Shadbolt}, \citenamefont {Matsuda},
  \citenamefont {Oguma}, \citenamefont {Itoh} \emph {et~al.}}]{carolanScience}%
  \BibitemOpen
  \bibfield  {author} {\bibinfo {author} {\bibfnamefont {J.}~\bibnamefont
  {Carolan}}, \bibinfo {author} {\bibfnamefont {C.}~\bibnamefont {Harrold}},
  \bibinfo {author} {\bibfnamefont {C.}~\bibnamefont {Sparrow}}, \bibinfo
  {author} {\bibfnamefont {E.}~\bibnamefont {Mart{\'\i}n-L{\'o}pez}}, \bibinfo
  {author} {\bibfnamefont {N.~J.}\ \bibnamefont {Russell}}, \bibinfo {author}
  {\bibfnamefont {J.~W.}\ \bibnamefont {Silverstone}}, \bibinfo {author}
  {\bibfnamefont {P.~J.}\ \bibnamefont {Shadbolt}}, \bibinfo {author}
  {\bibfnamefont {N.}~\bibnamefont {Matsuda}}, \bibinfo {author} {\bibfnamefont
  {M.}~\bibnamefont {Oguma}}, \bibinfo {author} {\bibfnamefont
  {M.}~\bibnamefont {Itoh}}, \emph {et~al.},\ }\bibfield  {title} {\bibinfo
  {title} {Universal linear optics},\ }\href@noop {} {\bibfield  {journal}
  {\bibinfo  {journal} {Science}\ }\textbf {\bibinfo {volume} {349}},\ \bibinfo
  {pages} {711} (\bibinfo {year} {2015})}\BibitemShut {NoStop}%
\bibitem [{\citenamefont {Reck}\ \emph {et~al.}(1994)\citenamefont {Reck},
  \citenamefont {Zeilinger}, \citenamefont {Bernstein},\ and\ \citenamefont
  {Bertani}}]{reck1994experimental}%
  \BibitemOpen
  \bibfield  {author} {\bibinfo {author} {\bibfnamefont {M.}~\bibnamefont
  {Reck}}, \bibinfo {author} {\bibfnamefont {A.}~\bibnamefont {Zeilinger}},
  \bibinfo {author} {\bibfnamefont {H.~J.}\ \bibnamefont {Bernstein}},\ and\
  \bibinfo {author} {\bibfnamefont {P.}~\bibnamefont {Bertani}},\ }\bibfield
  {title} {\bibinfo {title} {Experimental realization of any discrete unitary
  operator},\ }\href@noop {} {\bibfield  {journal} {\bibinfo  {journal}
  {Physical review letters}\ }\textbf {\bibinfo {volume} {73}},\ \bibinfo
  {pages} {58} (\bibinfo {year} {1994})}\BibitemShut {NoStop}%
\bibitem [{\citenamefont {Clements}\ \emph {et~al.}(2016)\citenamefont
  {Clements}, \citenamefont {Humphreys}, \citenamefont {Metcalf}, \citenamefont
  {Kolthammer},\ and\ \citenamefont {Walmsley}}]{clements2016optimal}%
  \BibitemOpen
  \bibfield  {author} {\bibinfo {author} {\bibfnamefont {W.~R.}\ \bibnamefont
  {Clements}}, \bibinfo {author} {\bibfnamefont {P.~C.}\ \bibnamefont
  {Humphreys}}, \bibinfo {author} {\bibfnamefont {B.~J.}\ \bibnamefont
  {Metcalf}}, \bibinfo {author} {\bibfnamefont {W.~S.}\ \bibnamefont
  {Kolthammer}},\ and\ \bibinfo {author} {\bibfnamefont {I.~A.}\ \bibnamefont
  {Walmsley}},\ }\bibfield  {title} {\bibinfo {title} {Optimal design for
  universal multiport interferometers},\ }\href@noop {} {\bibfield  {journal}
  {\bibinfo  {journal} {Optica}\ }\textbf {\bibinfo {volume} {3}},\ \bibinfo
  {pages} {1460} (\bibinfo {year} {2016})}\BibitemShut {NoStop}%
\bibitem [{\citenamefont {Chin}\ and\ \citenamefont
  {Huh}(2018)}]{chin2018generalized}%
  \BibitemOpen
  \bibfield  {author} {\bibinfo {author} {\bibfnamefont {S.}~\bibnamefont
  {Chin}}\ and\ \bibinfo {author} {\bibfnamefont {J.}~\bibnamefont {Huh}},\
  }\bibfield  {title} {\bibinfo {title} {Generalized concurrence in boson
  sampling},\ }\href@noop {} {\bibfield  {journal} {\bibinfo  {journal}
  {Scientific reports}\ }\textbf {\bibinfo {volume} {8}},\ \bibinfo {pages}
  {6101} (\bibinfo {year} {2018})}\BibitemShut {NoStop}%
\bibitem [{\citenamefont {Qi}\ \emph {et~al.}(2018)\citenamefont {Qi},
  \citenamefont {Helt}, \citenamefont {Su}, \citenamefont {Vernon},\ and\
  \citenamefont {Br{\'a}dler}}]{qi2018linear}%
  \BibitemOpen
  \bibfield  {author} {\bibinfo {author} {\bibfnamefont {H.}~\bibnamefont
  {Qi}}, \bibinfo {author} {\bibfnamefont {L.~G.}\ \bibnamefont {Helt}},
  \bibinfo {author} {\bibfnamefont {D.}~\bibnamefont {Su}}, \bibinfo {author}
  {\bibfnamefont {Z.}~\bibnamefont {Vernon}},\ and\ \bibinfo {author}
  {\bibfnamefont {K.}~\bibnamefont {Br{\'a}dler}},\ }\bibfield  {title}
  {\bibinfo {title} {Linear multiport photonic interferometers: loss analysis
  of temporally-encoded architectures},\ }\bibfield  {journal} {\bibinfo
  {journal} {arXiv preprint arXiv:1812.07015}\ }\href
  {https://doi.org/https://doi.org/10.48550/arXiv.1812.07015}
  {https://doi.org/10.48550/arXiv.1812.07015} (\bibinfo {year}
  {2018})\BibitemShut {NoStop}%
\bibitem [{\citenamefont {Ferrante}\ and\ \citenamefont
  {Frigo}(2012)}]{ferrante2012note}%
  \BibitemOpen
  \bibfield  {author} {\bibinfo {author} {\bibfnamefont {M.}~\bibnamefont
  {Ferrante}}\ and\ \bibinfo {author} {\bibfnamefont {N.}~\bibnamefont
  {Frigo}},\ }\bibfield  {title} {\bibinfo {title} {A note on the
  coupon-collector's problem with multiple arrivals and the random sampling},\
  }\href@noop {} {\bibfield  {journal} {\bibinfo  {journal} {arXiv preprint
  arXiv:1209.2667}\ } (\bibinfo {year} {2012})}\BibitemShut {NoStop}%
\bibitem [{\citenamefont {Ding}\ \emph {et~al.}(2016)\citenamefont {Ding},
  \citenamefont {He}, \citenamefont {Duan}, \citenamefont {Gregersen},
  \citenamefont {Chen}, \citenamefont {Unsleber}, \citenamefont {Maier},
  \citenamefont {Schneider}, \citenamefont {Kamp}, \citenamefont {H{\"o}fling}
  \emph {et~al.}}]{ding2016demand}%
  \BibitemOpen
  \bibfield  {author} {\bibinfo {author} {\bibfnamefont {X.}~\bibnamefont
  {Ding}}, \bibinfo {author} {\bibfnamefont {Y.}~\bibnamefont {He}}, \bibinfo
  {author} {\bibfnamefont {Z.-C.}\ \bibnamefont {Duan}}, \bibinfo {author}
  {\bibfnamefont {N.}~\bibnamefont {Gregersen}}, \bibinfo {author}
  {\bibfnamefont {M.-C.}\ \bibnamefont {Chen}}, \bibinfo {author}
  {\bibfnamefont {S.}~\bibnamefont {Unsleber}}, \bibinfo {author}
  {\bibfnamefont {S.}~\bibnamefont {Maier}}, \bibinfo {author} {\bibfnamefont
  {C.}~\bibnamefont {Schneider}}, \bibinfo {author} {\bibfnamefont
  {M.}~\bibnamefont {Kamp}}, \bibinfo {author} {\bibfnamefont {S.}~\bibnamefont
  {H{\"o}fling}}, \emph {et~al.},\ }\bibfield  {title} {\bibinfo {title}
  {On-demand single photons with high extraction efficiency and near-unity
  indistinguishability from a resonantly driven quantum dot in a micropillar},\
  }\href@noop {} {\bibfield  {journal} {\bibinfo  {journal} {Physical review
  letters}\ }\textbf {\bibinfo {volume} {116}},\ \bibinfo {pages} {020401}
  (\bibinfo {year} {2016})}\BibitemShut {NoStop}%
\bibitem [{\citenamefont {Dree{\ss}en}\ \emph {et~al.}(2018)\citenamefont
  {Dree{\ss}en}, \citenamefont {Ouellet-Plamondon}, \citenamefont {Tighineanu},
  \citenamefont {Zhou}, \citenamefont {Midolo}, \citenamefont {S{\o}rensen},\
  and\ \citenamefont {Lodahl}}]{Dree_en_2018}%
  \BibitemOpen
  \bibfield  {author} {\bibinfo {author} {\bibfnamefont {C.~L.}\ \bibnamefont
  {Dree{\ss}en}}, \bibinfo {author} {\bibfnamefont {C.}~\bibnamefont
  {Ouellet-Plamondon}}, \bibinfo {author} {\bibfnamefont {P.}~\bibnamefont
  {Tighineanu}}, \bibinfo {author} {\bibfnamefont {X.}~\bibnamefont {Zhou}},
  \bibinfo {author} {\bibfnamefont {L.}~\bibnamefont {Midolo}}, \bibinfo
  {author} {\bibfnamefont {A.~S.}\ \bibnamefont {S{\o}rensen}},\ and\ \bibinfo
  {author} {\bibfnamefont {P.}~\bibnamefont {Lodahl}},\ }\bibfield  {title}
  {\bibinfo {title} {Suppressing phonon decoherence of high performance
  single-photon sources in nanophotonic waveguides},\ }\href
  {https://doi.org/10.1088/2058-9565/aadbb8} {\bibfield  {journal} {\bibinfo
  {journal} {Quantum Science and Technology}\ }\textbf {\bibinfo {volume}
  {4}},\ \bibinfo {pages} {015003} (\bibinfo {year} {2018})}\BibitemShut
  {NoStop}%
\bibitem [{\citenamefont {Taballione}\ \emph {et~al.}(2023)\citenamefont
  {Taballione}, \citenamefont {Anguita}, \citenamefont {de~Goede},
  \citenamefont {Venderbosch}, \citenamefont {Kassenberg}, \citenamefont
  {Snijders}, \citenamefont {Kannan}, \citenamefont {Vleeshouwers},
  \citenamefont {Smith}, \citenamefont {Epping} \emph
  {et~al.}}]{taballione202320}%
  \BibitemOpen
  \bibfield  {author} {\bibinfo {author} {\bibfnamefont {C.}~\bibnamefont
  {Taballione}}, \bibinfo {author} {\bibfnamefont {M.~C.}\ \bibnamefont
  {Anguita}}, \bibinfo {author} {\bibfnamefont {M.}~\bibnamefont {de~Goede}},
  \bibinfo {author} {\bibfnamefont {P.}~\bibnamefont {Venderbosch}}, \bibinfo
  {author} {\bibfnamefont {B.}~\bibnamefont {Kassenberg}}, \bibinfo {author}
  {\bibfnamefont {H.}~\bibnamefont {Snijders}}, \bibinfo {author}
  {\bibfnamefont {N.}~\bibnamefont {Kannan}}, \bibinfo {author} {\bibfnamefont
  {W.~L.}\ \bibnamefont {Vleeshouwers}}, \bibinfo {author} {\bibfnamefont
  {D.}~\bibnamefont {Smith}}, \bibinfo {author} {\bibfnamefont {J.~P.}\
  \bibnamefont {Epping}}, \emph {et~al.},\ }\bibfield  {title} {\bibinfo
  {title} {20-mode universal quantum photonic processor},\ }\href@noop {}
  {\bibfield  {journal} {\bibinfo  {journal} {Quantum}\ }\textbf {\bibinfo
  {volume} {7}},\ \bibinfo {pages} {1071} (\bibinfo {year} {2023})}\BibitemShut
  {NoStop}%
\bibitem [{\citenamefont {Chen}\ \emph {et~al.}(2023)\citenamefont {Chen},
  \citenamefont {Peng}, \citenamefont {Guo}, \citenamefont {Gu}, \citenamefont
  {Ding}, \citenamefont {Liu}, \citenamefont {You}, \citenamefont {Qin},
  \citenamefont {Wang}, \citenamefont {He}, \citenamefont {Renema},
  \citenamefont {Huo}, \citenamefont {Wang}, \citenamefont {Lu},\ and\
  \citenamefont {Pan}}]{chen2023heralded}%
  \BibitemOpen
  \bibfield  {author} {\bibinfo {author} {\bibfnamefont {S.}~\bibnamefont
  {Chen}}, \bibinfo {author} {\bibfnamefont {L.-C.}\ \bibnamefont {Peng}},
  \bibinfo {author} {\bibfnamefont {Y.-P.}\ \bibnamefont {Guo}}, \bibinfo
  {author} {\bibfnamefont {X.-M.}\ \bibnamefont {Gu}}, \bibinfo {author}
  {\bibfnamefont {X.}~\bibnamefont {Ding}}, \bibinfo {author} {\bibfnamefont
  {R.-Z.}\ \bibnamefont {Liu}}, \bibinfo {author} {\bibfnamefont
  {X.}~\bibnamefont {You}}, \bibinfo {author} {\bibfnamefont {J.}~\bibnamefont
  {Qin}}, \bibinfo {author} {\bibfnamefont {Y.-F.}\ \bibnamefont {Wang}},
  \bibinfo {author} {\bibfnamefont {Y.-M.}\ \bibnamefont {He}}, \bibinfo
  {author} {\bibfnamefont {J.~J.}\ \bibnamefont {Renema}}, \bibinfo {author}
  {\bibfnamefont {Y.-H.}\ \bibnamefont {Huo}}, \bibinfo {author} {\bibfnamefont
  {H.}~\bibnamefont {Wang}}, \bibinfo {author} {\bibfnamefont {C.-Y.}\
  \bibnamefont {Lu}},\ and\ \bibinfo {author} {\bibfnamefont {J.-W.}\
  \bibnamefont {Pan}},\ }\href
  {https://doi.org/https://doi.org/10.48550/arXiv.2307.02189} {\bibinfo {title}
  {Heralded three-photon entanglement from a single-photon source on a photonic
  chip}} (\bibinfo {year} {2023}),\ \Eprint {https://arxiv.org/abs/2307.02189}
  {arXiv:2307.02189 [quant-ph]} \BibitemShut {NoStop}%
\bibitem [{\citenamefont {Wang}\ \emph {et~al.}(2023)\citenamefont {Wang},
  \citenamefont {Faurby}, \citenamefont {Ruf}, \citenamefont {Sund},
  \citenamefont {Nielsen}, \citenamefont {Volet}, \citenamefont {Heck},
  \citenamefont {Bart}, \citenamefont {Wieck}, \citenamefont {Ludwig},
  \citenamefont {Midolo}, \citenamefont {Paesani},\ and\ \citenamefont
  {Lodahl}}]{wang2023deterministic}%
  \BibitemOpen
  \bibfield  {author} {\bibinfo {author} {\bibfnamefont {Y.}~\bibnamefont
  {Wang}}, \bibinfo {author} {\bibfnamefont {C.~F.~D.}\ \bibnamefont {Faurby}},
  \bibinfo {author} {\bibfnamefont {F.}~\bibnamefont {Ruf}}, \bibinfo {author}
  {\bibfnamefont {P.~I.}\ \bibnamefont {Sund}}, \bibinfo {author}
  {\bibfnamefont {K.~H.}\ \bibnamefont {Nielsen}}, \bibinfo {author}
  {\bibfnamefont {N.}~\bibnamefont {Volet}}, \bibinfo {author} {\bibfnamefont
  {M.~J.~R.}\ \bibnamefont {Heck}}, \bibinfo {author} {\bibfnamefont
  {N.}~\bibnamefont {Bart}}, \bibinfo {author} {\bibfnamefont {A.~D.}\
  \bibnamefont {Wieck}}, \bibinfo {author} {\bibfnamefont {A.}~\bibnamefont
  {Ludwig}}, \bibinfo {author} {\bibfnamefont {L.}~\bibnamefont {Midolo}},
  \bibinfo {author} {\bibfnamefont {S.}~\bibnamefont {Paesani}},\ and\ \bibinfo
  {author} {\bibfnamefont {P.}~\bibnamefont {Lodahl}},\ }\href
  {https://doi.org/https://doi.org/10.48550/arXiv.2302.06282} {\bibinfo {title}
  {Deterministic photon source interfaced with a programmable silicon-nitride
  integrated circuit}} (\bibinfo {year} {2023}),\ \Eprint
  {https://arxiv.org/abs/2302.06282} {arXiv:2302.06282 [quant-ph]} \BibitemShut
  {NoStop}%
\bibitem [{\citenamefont {Pedersen}(2020)}]{frejaThesis}%
  \BibitemOpen
  \bibfield  {author} {\bibinfo {author} {\bibfnamefont {F.~T.}\ \bibnamefont
  {Pedersen}},\ }\emph {\bibinfo {title} {Deterministic Single and Multi-Photon
  Sources with Quantum dots in Planar Nanostructures}},\ \href@noop {}
  {\bibinfo {type} {Phd thesis}},\ \bibinfo  {school} {University of
  Copenhagen} (\bibinfo {year} {2020})\BibitemShut {NoStop}%
\bibitem [{\citenamefont {Tiecke}\ \emph {et~al.}(2015)\citenamefont {Tiecke},
  \citenamefont {Nayak}, \citenamefont {Thompson}, \citenamefont {Peyronel},
  \citenamefont {de~Leon}, \citenamefont {Vuleti{\'c}},\ and\ \citenamefont
  {Lukin}}]{tiecke2015efficient}%
  \BibitemOpen
  \bibfield  {author} {\bibinfo {author} {\bibfnamefont {T.}~\bibnamefont
  {Tiecke}}, \bibinfo {author} {\bibfnamefont {K.}~\bibnamefont {Nayak}},
  \bibinfo {author} {\bibfnamefont {J.~D.}\ \bibnamefont {Thompson}}, \bibinfo
  {author} {\bibfnamefont {T.}~\bibnamefont {Peyronel}}, \bibinfo {author}
  {\bibfnamefont {N.~P.}\ \bibnamefont {de~Leon}}, \bibinfo {author}
  {\bibfnamefont {V.}~\bibnamefont {Vuleti{\'c}}},\ and\ \bibinfo {author}
  {\bibfnamefont {M.}~\bibnamefont {Lukin}},\ }\bibfield  {title} {\bibinfo
  {title} {Efficient fiber-optical interface for nanophotonic devices},\
  }\href@noop {} {\bibfield  {journal} {\bibinfo  {journal} {Optica}\ }\textbf
  {\bibinfo {volume} {2}},\ \bibinfo {pages} {70} (\bibinfo {year}
  {2015})}\BibitemShut {NoStop}%
\bibitem [{\citenamefont {Notaros}\ \emph {et~al.}(2016)\citenamefont
  {Notaros}, \citenamefont {Pavanello}, \citenamefont {Wade}, \citenamefont
  {Gentry}, \citenamefont {Atabaki}, \citenamefont {Alloatti}, \citenamefont
  {Ram},\ and\ \citenamefont {Popovi{\'c}}}]{notaros2016ultra}%
  \BibitemOpen
  \bibfield  {author} {\bibinfo {author} {\bibfnamefont {J.}~\bibnamefont
  {Notaros}}, \bibinfo {author} {\bibfnamefont {F.}~\bibnamefont {Pavanello}},
  \bibinfo {author} {\bibfnamefont {M.~T.}\ \bibnamefont {Wade}}, \bibinfo
  {author} {\bibfnamefont {C.~M.}\ \bibnamefont {Gentry}}, \bibinfo {author}
  {\bibfnamefont {A.}~\bibnamefont {Atabaki}}, \bibinfo {author} {\bibfnamefont
  {L.}~\bibnamefont {Alloatti}}, \bibinfo {author} {\bibfnamefont {R.~J.}\
  \bibnamefont {Ram}},\ and\ \bibinfo {author} {\bibfnamefont {M.~A.}\
  \bibnamefont {Popovi{\'c}}},\ }\bibfield  {title} {\bibinfo {title}
  {Ultra-efficient cmos fiber-to-chip grating couplers},\ }in\ \href@noop {}
  {\emph {\bibinfo {booktitle} {2016 Optical Fiber Communications Conference
  and Exhibition (OFC)}}}\ (\bibinfo {organization} {IEEE},\ \bibinfo {year}
  {2016})\ pp.\ \bibinfo {pages} {1--3}\BibitemShut {NoStop}%
\bibitem [{\citenamefont {Marchetti}\ \emph {et~al.}(2019)\citenamefont
  {Marchetti}, \citenamefont {Lacava}, \citenamefont {Carroll}, \citenamefont
  {Gradkowski},\ and\ \citenamefont {Minzioni}}]{marchetti2019coupling}%
  \BibitemOpen
  \bibfield  {author} {\bibinfo {author} {\bibfnamefont {R.}~\bibnamefont
  {Marchetti}}, \bibinfo {author} {\bibfnamefont {C.}~\bibnamefont {Lacava}},
  \bibinfo {author} {\bibfnamefont {L.}~\bibnamefont {Carroll}}, \bibinfo
  {author} {\bibfnamefont {K.}~\bibnamefont {Gradkowski}},\ and\ \bibinfo
  {author} {\bibfnamefont {P.}~\bibnamefont {Minzioni}},\ }\bibfield  {title}
  {\bibinfo {title} {Coupling strategies for silicon photonics integrated
  chips},\ }\href@noop {} {\bibfield  {journal} {\bibinfo  {journal} {Photonics
  Research}\ }\textbf {\bibinfo {volume} {7}},\ \bibinfo {pages} {201}
  (\bibinfo {year} {2019})}\BibitemShut {NoStop}%
\end{thebibliography}%
\appendix
\section{On the computational complexity of boson sampling with collisions}
Here we will evaluate the computational complexity of boson sampling with collisions, which is relevant in the case of boson sampling with added lost photons where photon collisions and photon loss are indistinguishable due to the use of threshold detectors. 
To do so, we employ the approach used in Ref. \cite{chin2018generalized} which we summarize below. 
In this approach, the minimal computation time $\mathcal{T}_{\text{min}}(\Vec{n}, \Vec{m})$ is given as a function of the input state $\Vec{n}$ and output state $\Vec{m}$ in the Fock basis, where a state is defined as
\begin{align}
    \Vec{n} = (n_1, n_2, ..., n_M), \quad \sum_i n_i = N
\end{align}
where $n_i$ is the number of photons in the $i$th mode, $M$ is the total number of modes, and $N$ is the total number of photons.

The minimal computation time, $\mathcal{T}_{\text{min}}(\Vec{n}, \Vec{m})$ with input state $\Vec{n}$ and output state $\Vec{m}$ is found to be
\begin{align}
    \mathcal{T}_{\text{min}}(\Vec{n}, \Vec{m}) = \mathcal{O}\left(\min\left\{\sum_{k=0}^{\alpha_{\Vec{n}}} X_{k}(\Vec{n}), \sum_{l=0}^{\alpha_{\Vec{m}}} X_{l}(\Vec{m})\right\} \alpha_{\Vec{n}}\alpha_{\Vec{m}}\right).
\end{align}
Here, $\alpha_{\Vec{n}}$ is the number of nonzero elements in the Fock state vector $\Vec{n}$, called the Fock state coherence rank, and $X_{k}(\Vec{n})$ is the $k$-th elementary symmetric potential defined as
\begin{align}
    X_{k}(\Vec{n}) = \sum_{i_1<i_2<...<i_k=1} ^{\alpha_{\Vec{n}}} n_{i1} n_{i2} ... n_{ik}, (0\leq k \leq \alpha_{\Vec{n}} \leq M).
    \label{eq:elementary_symmetric_potential}
\end{align}

The minimal computation time is then determined by the quantity $\sum_{k=0}^{\alpha_{\Vec{n}}} X_{k}(\Vec{n})$ for the input state and the output state. 
The input state will be collision-free, $\Vec{n}_{\text{collision-free}}$ for which the following holds \cite{chin2018generalized}
\begin{align}
    \sum_{k=0}^{\alpha_{\Vec{n}}} X_{k}(\Vec{n}_{\text{collision-free}}) = 2^{N} = 2^{\alpha_{\Vec{n}}}.
\end{align}

Adding a collision to a collision-free state corresponds to increasing one of the nonzero numbers $n_i$ in $\Vec{n}$. From the definition of Eq. \eqref{eq:elementary_symmetric_potential}, we can see that the value of $\sum_{k=0}^{\alpha_{\Vec{n}}} X_{k}(\Vec{n})$ then has to increase. As such, the following inequality will be true for any state $\Vec{n}$
\begin{align}
    \sum_{k=0}^{\alpha_{\Vec{n}}} X_{k}(\Vec{n}) \geq 2^{\alpha_{\Vec{n}}}.
\end{align}

Thus we can say that the minimal computation time has to scale at least as 
$\mathcal{O}(2^{\alpha_{\Vec{m}}}\alpha_{\Vec{n}}\alpha_{\Vec{m}})$.
Accordingly, we can conclude that the computational complexity of boson sampling with collisions is not diminished compared to the computational complexity of collision-free boson sampling with the same number of nonzero elements.

\section{The size of the Hilbert subspace for a given number of photon collisions.}
The number of basis states in the Hilbert space with a given number of collisions is equal to the product of the number of ways one can distribute $d$ nonzero modes in $m$ modes, and the number of ways one can place $p-l-d$ collisions into $d$ nonzero modes. The first number is equal to the number of combinations without replacements with $d$ choices from $m$ possibilities:
\begin{align*}
    \binom{m}{d}.
\end{align*}
The second number is equal to the number of combinations with replacements with $p-l-d$ choices from $d$ possibilities
\begin{align*}
    &\binom{(p-l-d) + d - 1}{p-l-d} \\ 
    =&\binom{p-l-1}{p-l-d}
\end{align*}
Consequently the size of the Hilbert subspace with $p-d-l$ collisions, $n_{\text{collisions}}(p, d, l, m)$, will be
\begin{align}
    n_{\text{collisions}}(p, d, l, m) = \binom{m}{d} \cdot \binom{p-l-1}{p-l-d}.
    \label{eq:collisional_size}
\end{align}

The size of the full Hilbert space, $n_{\text{full}}$ will be the number of ways one can place $p-l$ photons into $m$ modes. This is equivalent to the number of combinations with replacements of $p-l$ choices with $m$ possibilities:
\begin{align}
    n_{\text{full}}(m, p, l) = \binom{m + p-l - 1}{p-l}. 
    \label{eq:FullHilbert}
\end{align}

We can then take the ratio between Eqs. \eqref{eq:collisional_size} and \eqref{eq:FullHilbert} to find Eq. \eqref{eq:effective_post_selection_efficiency}

\section{The downsides of using loop architectures for time-bin interferometer }
Though it is possible to implement large multimode interferometers using only a single physical MZI connected to fiber delay loops, as in Refs. \cite{motes2014scalable, he2017time}, this has two major downsides: higher propagation loss and severely reduced input state rate. Much in the same way as in the cascaded case, a column of MZIs in the Clements or Rectangular scheme can be implemented by sending the time-bins through the MZI one by one and reconfiguring the MZI transformation for each time-bin. In order to reuse the same physical MZI to implement additional columns, we can connect the outputs to the inputs through a delay loop, where the delay is sufficiently long that all output modes back at the input after the previous MZI column has been finished. As each column processes up to $m/2$ time-bins, this requires that the loops have a delay of at least $m/2$ time-bins. This is in comparison to the cascaded scheme where no such delay is necessary apart from the delay of one time-bin in one of the modes, which will still be present in the loop architecture. To see the difference, we can compare the total delay, $t_{\text{delay}}$ for the worst case of the Clements scheme with a cascaded architecture and a loop architecture
\begin{align}
    t_{\text{delay, cascaded}} = \frac{m}{2}-1 \cdot \tau, \\
    t_{\text{delay, loop}} = (m-1)(\frac{m}{2}-1) \cdot \tau,
    \label{eq:delay loop}
\end{align}
where $\tau$ corresponds to the separation between time-bins. In other words, the total propagation loss scales linearly with the number of modes for cascaded time-bin interferometers, whereas it scales quadratically with the number of modes for loop time-bin interferometers.

The second downside is that one has to wait for the full output state to come out of the interferometer before processing a new input state. The time difference between the first time-bin in the input state and the last time-bin of the output state is equivalent to the delay in Eq. \eqref{eq:delay loop}. As the time-bin separation is related to the rate of the single-photon source, the input generation rate, $r_{\text{input}}$ for the loop time-bin interferometer will be given by
\begin{align}
    r_{\text{input}} = \frac{2r_{\text{single-photon}}}{m(m-1))}.
\end{align}
This is approximately a factor $1/m$ worse than the corresponding rate for a cascaded interferometer.

\end{document}